\documentclass[aps,pre,reprint,showpacs,amsfonts,amssymb,amsmath,longbibliography]{revtex4-1}

\usepackage{bm}
\usepackage{graphicx}
\usepackage[caption=false]{subfig}
\usepackage[pdftex]{hyperref}
\begin{document}
%
%
%
% ==================================================================================================== 
% HEADER
% ====================================================================================================
%
\title{General and exact approach to percolation on random graphs}
\author{Antoine \surname{Allard}}
\altaffiliation[Now at ]{Departament de F\'isica Fonamental, Universitat de Barcelona, Carrer de Mart\'i i Franqu\`es 1, 08028 Barcelona, Spain}
\author{Laurent \surname{H\'ebert-Dufresne}}
\altaffiliation[Now at ]{Santa Fe Institute, Santa Fe, New Mexico 87501, USA}
\author{Jean-Gabriel \surname{Young}}
\author{Louis J. \surname{Dub\'e}}
\affiliation{D\'epartement de physique, de g\'enie physique, et d'optique, Universit\'e Laval, Qu\'ebec (Qc), Canada G1V 0A6}
\date{\today}
\begin{abstract}
  We present a comprehensive and versatile theoretical framework to study site and bond percolation on clustered and correlated random graphs. Our contribution can be summarized in three main points. (i) We introduce a set of iterative equations that solve the exact distribution of the size and composition of components in finite size quenched or random multitype graphs. (ii) We define a very general random graph ensemble that encompasses most of the models published to this day, and also that permits to model structural properties not yet included in a theoretical framework. Site and bond percolation on this ensemble is solved exactly in the infinite size limit using probability generating functions [i.e., the percolation threshold, the size and the composition of the giant (extensive) and small components]. Several examples and applications are also provided. (iii) Our approach can be adapted to model interdependent graphs---whose most striking feature is the emergence of an extensive component via a discontinuous phase transition---in an equally general fashion. We show how a graph can successively undergo a continuous then a discontinuous phase transition, and preliminary results suggest that clustering increases the amplitude of the discontinuity at the transition.
\end{abstract}
\maketitle
%
%
%
%
%
% ====================================================================================================
\section{Introduction}
% ====================================================================================================
%
Percolation on graphs offers a simple theoretical framework to model and investigate the behavior of many complex systems; noteworthy examples being the growth and the robustness of their structure \cite{Cohen10_ComplexNetworks,Dorogovtsev08_RevModPhys}, their observability \cite{Allard2014,Yang12_PhysRevLett}, as well as the effect of their structure on the propagation of emerging infectious agents \cite{Hebert-Dufresne13_SciRep,Meyers07_BullAmerMathSoc}. On the analytical front, recent progress has been mainly achieved within the Configuration Model (CM) paradigm \cite{Newman2010}, which, in the limit of large graphs, allows an exact and simple analytical treatment with the use of probability generating functions (pgf) \cite{Callaway00_PhysRevLett,Newman01_PhysRevE}. The versatility of the pgf method has triggered the development of many variants of the CM reproducing, to some extent, correlations and clustering found in real complex systems \cite{Allard12_JPhysA,Allard09_PhysRevE,Berchenko09_PhysRevLett,Ghoshal09_PhysRevE,Gleeson09_PhysRevE,Hackett11_IntJCompSystSci,Hebert-Dufresne13_PhysRevE,Karrer10_PhysRevE,Kenah07_PhysRevE,Leicht09_arXiv,Meyers06_JTheorBiol,Miller09_PhysRevE,Newman02_PhysRevLett,Newman03a_PhysRevE,Newman03b_PhysRevE,Newman09_PhysRevLett,Noel09_PhysRevE,Serrano06_PhysRevLett,Shi07_PhysicaA,Vazquez06_PhysRevE,Vazquez03_PhysRevE,Zlatic12_EPL}.

To move beyond what has been done thus far, we introduce a very general and comprehensive class of random graphs that increases significantly the nontrivial correlations and clustering patterns that can be handled analytically. Correlations and clustering are incorporated into the graphs through the use of types of vertices and types of stubs (i.e., half-edge stemming from vertices). Hence, by explicitly controlling \textit{who is connected to whom} and \textit{through what kind of connection}, our approach reproduces any correlations as long as they can be mapped unto this multitype framework. For instance, the type of the vertices can correspond to their degree (the number of neighbors) \cite{Newman02_PhysRevLett,Vazquez03_PhysRevE}, to their intrinsic properties such as age or ethnicity \cite{Allard09_PhysRevE,Newman03a_PhysRevE}, or to their position in the k-core structure of the graph \cite{Hebert-Dufresne13_PhysRevE}.

Furthermore the use of types of stubs explicitly accounts for different categories of connections. On the one hand, these differences may be of a \textit{conceptual} nature \cite{Kivela2014}. For instance in \textit{multilayer} or \textit{multiplex} graphs the type of an edge refers to the layer of interaction to which it belongs (e.g., family ties and acquaintances in social networks). On the other hand, the different types of stubs can describe different \textit{topological} functions. Since some edges may be undirected or directed, different types of stubs can be used to identify \textit{in-degrees}, \textit{out-degrees} or \textit{undirected degrees} \cite{Meyers06_JTheorBiol,Newman01_PhysRevE}. More importantly perhaps, stubs can be matched in groups of more than two vertices to form \textit{motifs}, also called \textit{hyperedges} \cite{Allard12_JPhysA,Karrer10_PhysRevE}, permitting the inclusion of clustering in a very general and natural fashion. These motifs can take a wide variety of forms: simple triangles, cliques of several hundreds of vertices, or arbitrary graphs with directed and multiple edges [see Fig.~\ref{fig:omnibus_allard_fig_1}\protect\subref{fig:omnibus_allard_fig_1a}]. Additionally, these motifs can have a quenched (i.e., fixed) or a random structure (e.g., multitype Erd\H{o}s-R\'enyi graphs).

We have developed a mathematical framework that solves the site \textit{and} bond percolation (hereafter \textit{hybrid} percolation) on this general class of random graphs. We build upon the well-known pgf-based formalism and obtain the analytical expression for the size of the extensive ``giant'' component, the percolation threshold, as well as the distribution of the size of the ``small'' components in the limit of large graph size. However, the pgf approach \textit{de facto} assumes locally tree-like graphs forbidding closed loops and therefore any clustering whatsoever. To circumvent this limitation, we present a set of iterative equations that exactly solves the size distribution of components in finite-size arbitrary, or quenched, graphs. These equations map the possible outcomes of hybrid percolation on any motifs (i.e., the size distribution of the components) unto a distribution of branching trees, and thereby reconcile the presence of motifs with the tree-like requirement of the pgf approach.

The general nature of our model acts as a \textit{theoretical laboratory} where the effect of a wide selection of structural features on the outcomes of hybrid percolation can be investigated on a common ground. To facilitate understanding and to provide support for our claims, several examples accompany the analysis and illustrate its practical implementation. Moreover, our model encompasses most variants of the CM published to date, we provide several examples supporting this claim as well.

Finally, we show how our approach can be adapted to model interdependent graphs---in which the extensive component emerges via a discontinuous transition instead of a continuous one \cite{Buldyrev10_Nature,Parshani2010}--- through a suitable change in the definition of what constitutes an extensive component (i.e., the order parameter). This adaptation show that a graph can successively undergo a continuous then a discontinuous phase transition \footnote{It has been brought to our attention after the completion of this manuscript that similar phenomena have been studied in Refs.~\cite{Bianconi2014,Wu2014a}.}, and provide a quantitative measure of the effect of clustering on the emergence of the extensive component.

The paper is organized as follows. In Sec.~\ref{sec:omnibus_small_graphs}, we introduce the set of iterative equations that exactly solves the size distribution of components in finite-size arbitrary graphs. In Sec.~\ref{sec:omnibus_pgf_formalism}, we formally define the general graph ensemble discussed above and obtain its exact structural properties under hybrid percolation (i.e., the size and composition of the components and the position of the percolation threshold). We then illustrate the workings of our formalism with several examples and special cases in Sec.~\ref{sec:omnibus_applications}. We finally show how our approach can be adapted to model interdependent graphs in Sec.~\ref{sec:omnibus_discontinuous}. Conclusions and final remarks are collected in Sec.~\ref{sec:omnibus_conclusion}.
%
%
%
%
%
% =================================================================================================
\section{Percolation on finite-size arbitrary graphs} \label{sec:omnibus_small_graphs}
% =================================================================================================
%
To reconcile the tree-like assumption of the pgf approach with the presence of motifs in graphs, the outcomes of percolation on these motifs---the distribution of the number of vertices that can be reached from a given vertex---must be obtained beforehand. These distributions can be computed by hand by enumerating each possible configuration where vertices and edges exist with given probabilities \cite{Karrer10_PhysRevE,Miller09_PhysRevE}. However, this procedure becomes rapidly unwieldly for motifs of more than a handful of vertices. Instead, we generalize the equations presented in Ref.~\cite{Allard12_EPL} to obtain a set of iterative equations that solve the outcome of hybrid percolation on small arbitrary graphs.
%
%
%
% =================================================================================================
\subsection{Multitype Erd\H{o}s-R\'enyi graphs} \label{sec:omnibus_small_graphs_multitype}
% =================================================================================================
%
\begin{figure*}[tb]
  \centering
  \subfloat[]{\label{fig:omnibus_allard_fig_1a} \includegraphics[width = 0.3\linewidth]{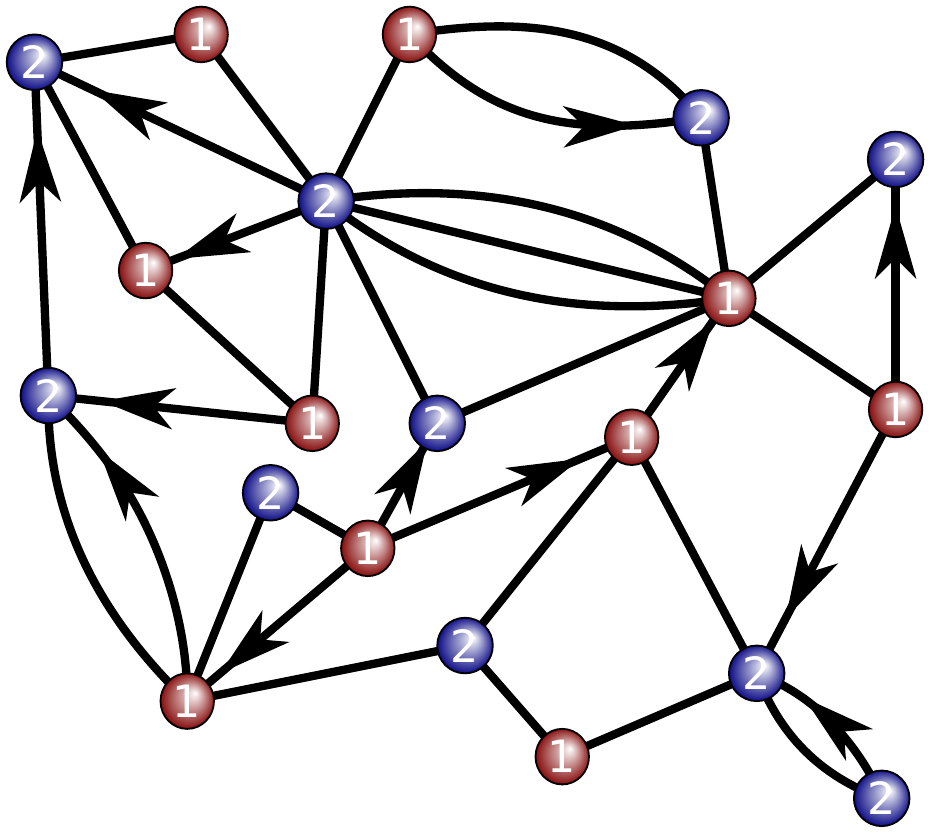}}
  \hspace{0.1\linewidth}
  \subfloat[]{\label{fig:omnibus_allard_fig_1b} \includegraphics[width = 0.45\linewidth]{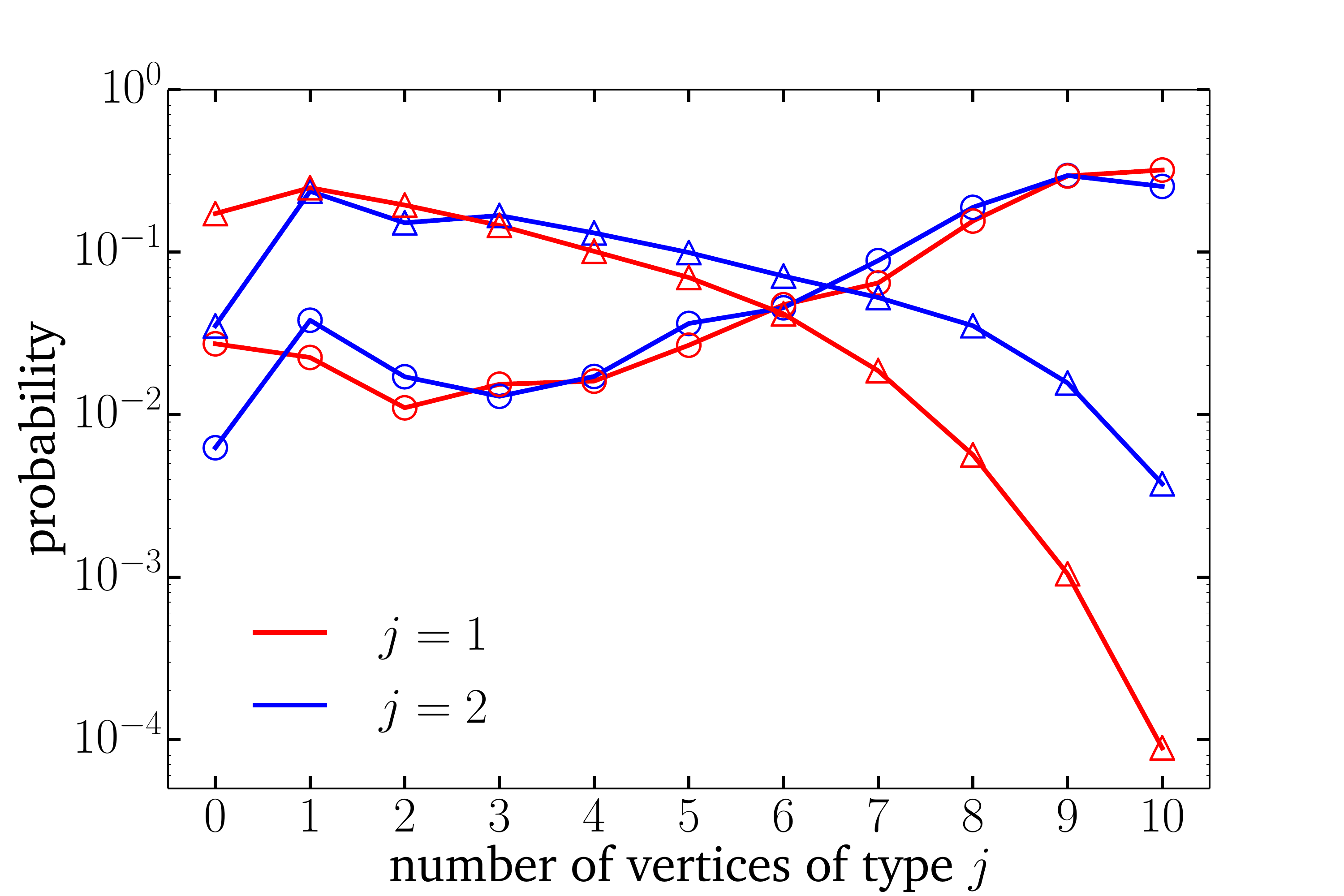}} \\
  \caption{\label{fig:omnibus_allard_fig_1}(Color online) \protect\subref{fig:omnibus_allard_fig_1a} Example of an arbitrary graph that can be handled by our framework. Blue and red represent vertex types 1 and 2. There are 10 vertices of each type. \protect\subref{fig:omnibus_allard_fig_1b} Distribution of the number of vertices of type $j$ in components reached from an initial vertex chosen at random for the graph shown in \protect\subref{fig:omnibus_allard_fig_1a}. Symbols are the results of $2.5\times10^8$ numerical simulations, and lines are the predictions of Eqs.~(\ref{eq:omnibus_gnp_1})--(\ref{eq:omnibus_gnp_3}). The distributions are discrete; lines have been added to guide the eye. Triangles ($\triangle$) correspond to pure site percolation with $\{\tilde{r}^\prime_s\} \equiv \{\tilde{r}^\prime_1,\tilde{r}^\prime_2\} = \{0.40,0.70\}$ and $\{\tilde{p}^\prime_{ij}\} \equiv \{\tilde{p}^\prime_{11},\tilde{p}^\prime_{12},\tilde{p}^\prime_{21},\tilde{p}^\prime_{22}\} = \{1.00,1.00,1.00,1.00\}$. Circles ($\circ$) correspond to hybrid percolation (site and bond) with $\{\tilde{r}^\prime_s\} \equiv \{\tilde{r}^\prime_1,\tilde{r}^\prime_2\} = \{0.95,0.90\}$ and $\{\tilde{p}^\prime_{ij}\} \equiv \{\tilde{p}^\prime_{11},\tilde{p}^\prime_{12},\tilde{p}^\prime_{21},\tilde{p}^\prime_{22}\} = \{1.00,0.90,0.85,0.95\}$. These probabilities are given in terms of the original vertex types to lighten the presentation (thus the use of a prime). To use the mapping described in Sec.~\ref{sec:omnibus_small_graphs_arbitrary}, a probability for each individual vertex and each individual edge must be defined. For instance, if vertex 8 is of type 1, we set $r_8=\tilde{r}^\prime_1$. Similarly, if vertex 5 is of type 2 and shares three undirected edges with vertex 8, we set $p_{58}=1-(1-\tilde{p}^\prime_{21})^3$ and $p_{85}=1-(1-\tilde{p}^\prime_{12})^3$.}
\end{figure*}
Let us first consider multitype random graphs as a generalization of the $\mathcal{G}_{n,p}$ model (i.e., Erd\H{o}s-R\'enyi random graphs) in which $n$ vertices are linked by edges that exist individually and independently with a probability $p$ \cite{Gilbert1959}. We generalize this model by labeling vertices using types; the set of types is noted $\mathcal{N}$, and there are a total of $|\mathcal{N}|$ types of vertices. A directed edge from a vertex of type $i$ towards a vertex of type $j$ (noted $i \rightarrow j$) exists with a probability $p_{ij}$ independently of other potential edges \footnote{An undirected edge between a type $i$ vertex and a type $j$ vertex (noted $i \leftrightarrow j$) therefore occurs with a probability $p_{ij}p_{ji}$. The symmetric case where $p_{ij}=p_{ji}$ for all $i$ and $j$ is statistically equivalent to the situation where all edges are undirected.}. For the sake of conciseness, we will refer to a graph composed of $n_i$ vertices of type $i$ (with $i=1,\ldots,|\mathcal{N}|$) with the vector $\bm{n}\equiv(n_1,\ldots,n_{|\mathcal{N}|})^\mathsf{T}$. We will use a similar notation for other quantities throughout this paper, unless specified otherwise.

Since edges may be directed, we define a component as the vertices that are reachable from a given initial vertex, including itself (i.e., the out-component rooted to this given vertex). This initial vertex is identified solely by its type since vertices of a given type are indistinguishable. We define $W_i(\bm{l}|\bm{n})$ as the probability that $\bm{l}\equiv(l_1,\ldots,l_{|\mathcal{N}|})^\mathsf{T}$ vertices can be reached from an initial vertex of type $i$ in a graph containing $\bm{n}$ vertices. The calculation of $W_i(\bm{l}|\bm{n})$ begins with the initial condition $W_i(\bm{\delta_i}|\bm{\delta_i}) = 1$, where $\bm{\delta_i}$ is the vector of Kronecker deltas $(\delta_{i1},\ldots,\delta_{i|\mathcal{N}|})^\mathsf{T}$ and corresponds to a single vertex of type $i$. This initial condition simply states that the probability of finding a component of one vertex of type $i$ in a graph containing one vertex of type $i$ is 1. Now suppose that there are other vertices in the graph and that it contains $\bm{n}$ vertices instead. The probability of finding a component of only one vertex of type $i$ in this graph, $W_i(\bm{\delta_i}|\bm{n})$, is equal to the probability that there is a component containing $\bm{\delta_i}$ vertex, $W_i(\bm{\delta_i}|\bm{\delta_i})$, (which in this case is equal to one) multiplied with the probability that none of the potential edges from the vertex in the component (i.e., the vertex of type $i$) towards the other vertices of the graph of size $\bm{n}$ exist
\begin{align} \label{eq:omnibus_gnp_prelim_1}
  W_i(\bm{\delta_i}|\bm{n})
    & = W_i(\bm{\delta_i}|\bm{\delta_i}) \prod_{k\in\mathcal{N}} (1-p_{ik})^{n_k-\delta_{ik}} \ .
\end{align} 
Let us now consider a component made of 2 vertices of type $i$, that we note $2\bm{\delta_i}$. By definition of the multitype random graphs, we know that $W_i(2\bm{\delta_i}|2\bm{\delta_i}) = p_{ii}$ since the component exists only if there is a directed edge from the initial vertex to the other vertex of type $i$. Following the steps leading to Eq.~\eqref{eq:omnibus_gnp_prelim_1}, the probability of finding a component of size $2\bm{\delta_i}$ from an initial vertex of type $i$ in a graph containing $\bm{n}$ vertices (we assume that $n_i \geq 2$) is
\begin{align} \label{eq:omnibus_gnp_prelim_2}
  W_i(2\bm{\delta_i}|\bm{n}) & = W_i(2\bm{\delta_i}|2\bm{\delta_i}) (n_i-1) \prod_{k\in\mathcal{N}} (1-p_{ik})^{2(n_k-2\delta_{ik})} \ ,
\end{align}
where the extra factor $(n_i-1)$ accounts for the number of ways to choose the second vertex among the $n_i-1$ available vertices of type $i$ in the graph, and $2(n_k-2\delta_{ik})$ is the number of potential edges from the two vertices of type $i$ towards the $n_k-2\delta_{ik}$ vertices of type $k$. Repeating this exercise for larger components, we obtain the following general form for a generic component of size $\bm{l}$
\begin{subequations} \label{eq:omnibus_gnp}
\begin{align} \label{eq:omnibus_gnp_1}
 W_i(\bm{l}|\bm{n}) = W_i(\bm{l}|\bm{l}) \prod_{j\in\mathcal{N}} \binom{n_j-\delta_{ij}}{l_j-\delta_{ij}} \prod_{k\in\mathcal{N}} (1-p_{jk})^{l_j(n_k-l_k)} \ ,
\end{align}
where $W_i(\bm{l}|\bm{l})$ is the probability that $\bm{l}$ vertices form a component considering an initial vertex of type $i$. In this last equation, the binomial coefficients count the number of ways the other $\bm{l}-\bm{\delta_i}$ vertices in the component can be chosen from the $\bm{n}-\bm{\delta_i}$ vertices available in the graph, and the other terms correspond to the probability that no other vertices can be reached from the $\bm{l}$ vertices in the component.

The only missing information in Eq.~\eqref{eq:omnibus_gnp_1} is the probability $W_i(\bm{l}|\bm{l})$. As seen in the two simple examples above, it is possible to compute the probability $W_i(\bm{l}|\bm{l})$ by hand, but this calculation becomes rapidly tedious as the size of the component increases. Fortunately, we can use Eq.~\eqref{eq:omnibus_gnp_1} to circumvent this difficulty. For example, from $W_i(\bm{\delta_i}|\bm{\delta_i}) = 1$, Eq.~\eqref{eq:omnibus_gnp_prelim_1} yields $W_i(\bm{\delta_i}|2\bm{\delta_i})=1-p_{ii}$. Since the probabilities must sum to 1 for a given graph size, we conclude that $W_i(2\bm{\delta_i}|2\bm{\delta_i})=1-W_i(\bm{\delta_i}|2\bm{\delta_i}) = p_{ii}$. Hence it is possible to build upon the probabilities computed for smaller graph size to obtain the missing probability $W_i(\bm{l}|\bm{l})$ by simply asking for normalization. In general terms,
\begin{align} \label{eq:omnibus_gnp_2}
 W_{i}(\bm{l}|\bm{l}) = 1 - \sum_{\bm{m}<\bm{l}} W_{i}(\bm{m}|\bm{l}) \ ,
\end{align}
where the probabilities $W_{i}(\bm{m}|\bm{l})$ are obtained with Eq.~\eqref{eq:omnibus_gnp_1}, and where the sum covers every possible instances of $\bm{m}$ such that $m_j \leq l_j$ for all $j$ but excludes the case in which all elements of the two vectors are equal (i.e., $m_j=l_j$ for every $j$). In short, Eqs.~(\ref{eq:omnibus_gnp_1})--(\ref{eq:omnibus_gnp_2}) are mutually dependent: the left-hand side of one feeds the right-hand side of the other. Thus, from a graph consisting of a single vertex (the initial condition), Eqs.~(\ref{eq:omnibus_gnp_1})--(\ref{eq:omnibus_gnp_2}) extend the graph to the desired size $\bm{n}$, and keep track of the component size distribution along the way to build the final distribution $\{W_i(\bm{l}|\bm{n})\}$.

A mass of information is produced during the iteration of Eqs.~(\ref{eq:omnibus_gnp_1})--(\ref{eq:omnibus_gnp_2}): the probability of finding every possible components $\bm{l}$ in each intermediate graph whose size is smaller than $\bm{n}$. When interested in bond percolation solely (as in Ref.~\cite{Allard12_EPL}), the only probabilities of interest are the ones related to the graph of maximum size $\bm{n}$. This ultimately leaves most of the calculated probabilities unused. However, if interested in hybrid percolation, that is when edges \textit{and} vertices exist with given probabilities, all the calculated probabilities can be put to contribution.% by weighting them by the proper probability. 

The probability for a graph of original size $\bm{n}$ to be left with $\bm{b}$ vertices after each of its vertices has been independently \textit{kept} with probabilities $\{r_j\}_{j\in\mathcal{N}}$ (i.e., a vertex of type $j$ is kept with probability $r_j$) is
\begin{align}
  B_i(\bm{b}|\bm{n}) \equiv \prod_{j\in\mathcal{N}} \binom{n_j-\delta_{ij}}{b_j-\delta_{ij}} r_j^{b_j-\delta_{ij}} (1-r_j)^{n_j-b_j} \ ,
\end{align}
where we assume that the initial vertex of type $i$ exists. Hence, from a starting vertex of type $i$, the probability to find a component of size $\bm{l}$ in a graph of original size $\bm{n}$ when vertices and edges exist with given probabilities, $Q_{i}(\bm{l}|\bm{n})$, is
\begin{align} \label{eq:omnibus_gnp_3}
 Q_{i}(\bm{l}|\bm{n}) & = \sum_{\bm{b}=\bm{l}}^{\bm{n}} W_i(\bm{l}|\bm{b}) B_i(\bm{b}|\bm{n}) \ ,
\end{align}
\end{subequations}
where the sum covers every possible instances of $\bm{b}$ such that $l_j \leq b_j \leq n_j$ for every $j\in\mathcal{N}$. Thus, by slightly increasing the computational effort, it is possible to incorporate site percolation into the systematic method introduced in Ref.~\cite{Allard12_EPL} for bond percolation. 
%
%
%
% =================================================================================================
\subsection{Arbitrary graphs} \label{sec:omnibus_small_graphs_arbitrary}
% =================================================================================================
%
Our framework can also be used to predict the outcomes of hybrid percolation on small arbitrary graphs. By small arbitrary graphs, we mean graphs of finite size with a fixed structure, in which edges may be directed and/or multiple, and whose vertices may belong to types. We use the \textit{adjacency} matrix $\mathbf{A}$, whose element $A_{ij}$ is the number of edges leaving vertex $i$ towards vertex $j$, to specify the structure. Figure~\ref{fig:omnibus_allard_fig_1}\protect\subref{fig:omnibus_allard_fig_1a} depicts an example of such an arbitrary graph. In such graphs, percolation corresponds to the random removal of edges and vertices according to some given probabilities which may depend on the type of the vertices involved. Predicting the outcome of percolation then consists in predicting the probability that a component of size $\bm{l}$ can be reached from a given initial vertex in a graph of size $\bm{n}$.

For Eqs.~\eqref{eq:omnibus_gnp_1}--\eqref{eq:omnibus_gnp_3} to be applicable, we need to map arbitrary graphs unto multitype random graphs. This mapping is achieved by assigning to each vertex its own type ($|\mathcal{N}|$ equals the number of vertices), and by setting the probabilities $\{p_{ij}\}$ to mimic the structure of the original arbitrary graph. To account for the fact that more than one edge may exist between two vertices in the original graph, we set $p_{ij} = 1 - (1-\tilde{p}_{ij})^{A_{ij}}$, where $\tilde{p}_{ij}$ is the probability that an individual edge from vertex $i$ to vertex $j$ remains after the random removal of edges in the arbitrary graph. Note that $\tilde{p}_{ij}$ may depend on the original types of vertices $i$ and $j$ in the graph [e.g., there are two types of vertices in Fig.~\ref{fig:omnibus_allard_fig_1}\protect\subref{fig:omnibus_allard_fig_1a}]. The same applies for the existence probabilities of vertices (i.e., $\{r_i\}$ must be equal to $\{\tilde{r}_i\}$). An example is given in the caption of Fig.~\ref{fig:omnibus_allard_fig_1}. Using this mapping, Eqs.~\eqref{eq:omnibus_gnp_1}--\eqref{eq:omnibus_gnp_3} offer a systematic procedure to compute the outcomes of hybrid percolation on small arbitrary graphs. Figure~\ref{fig:omnibus_allard_fig_1}\protect\subref{fig:omnibus_allard_fig_1b} compares the predictions of Eqs.~\eqref{eq:omnibus_gnp_1}--\eqref{eq:omnibus_gnp_3} with the results of numerical simulations for the arbitrary graph shown in Fig.~\ref{fig:omnibus_allard_fig_1}\protect\subref{fig:omnibus_allard_fig_1a}. As expected, a perfect agreement is observed.
%
%
%
%
%
% =================================================================================================
\section{Percolation on correlated and clustered infinite random graphs} \label{sec:omnibus_pgf_formalism}
% =================================================================================================
%
We now turn our attention to the generalized version of the CM briefly described in the Introduction. We provide a formal definition of the model, and analytically solve percolation for this general ensemble of random graphs.
%
%
%
% =================================================================================================
\subsection{A \textit{stub matching} scheme} \label{sec:omnibus_stub_matching}
% =================================================================================================
%
\begin{figure*}[tb]
  \centering
  \subfloat[]{\label{fig:omnibus_example_a} \includegraphics[width = 0.35\linewidth]{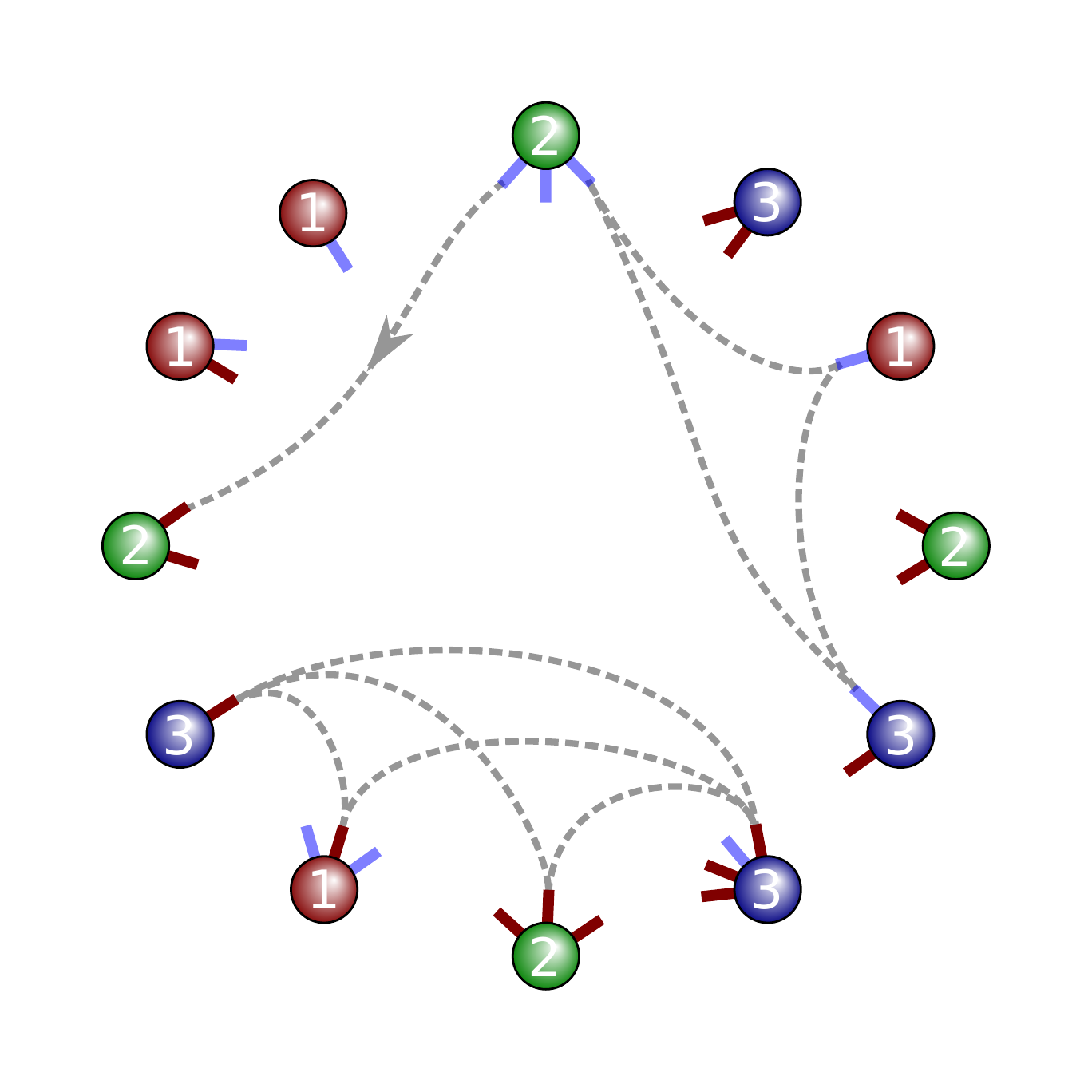}}
  \hspace{0.1\linewidth}
  \subfloat[]{\label{fig:omnibus_example_b} \includegraphics[width = 0.4\linewidth]{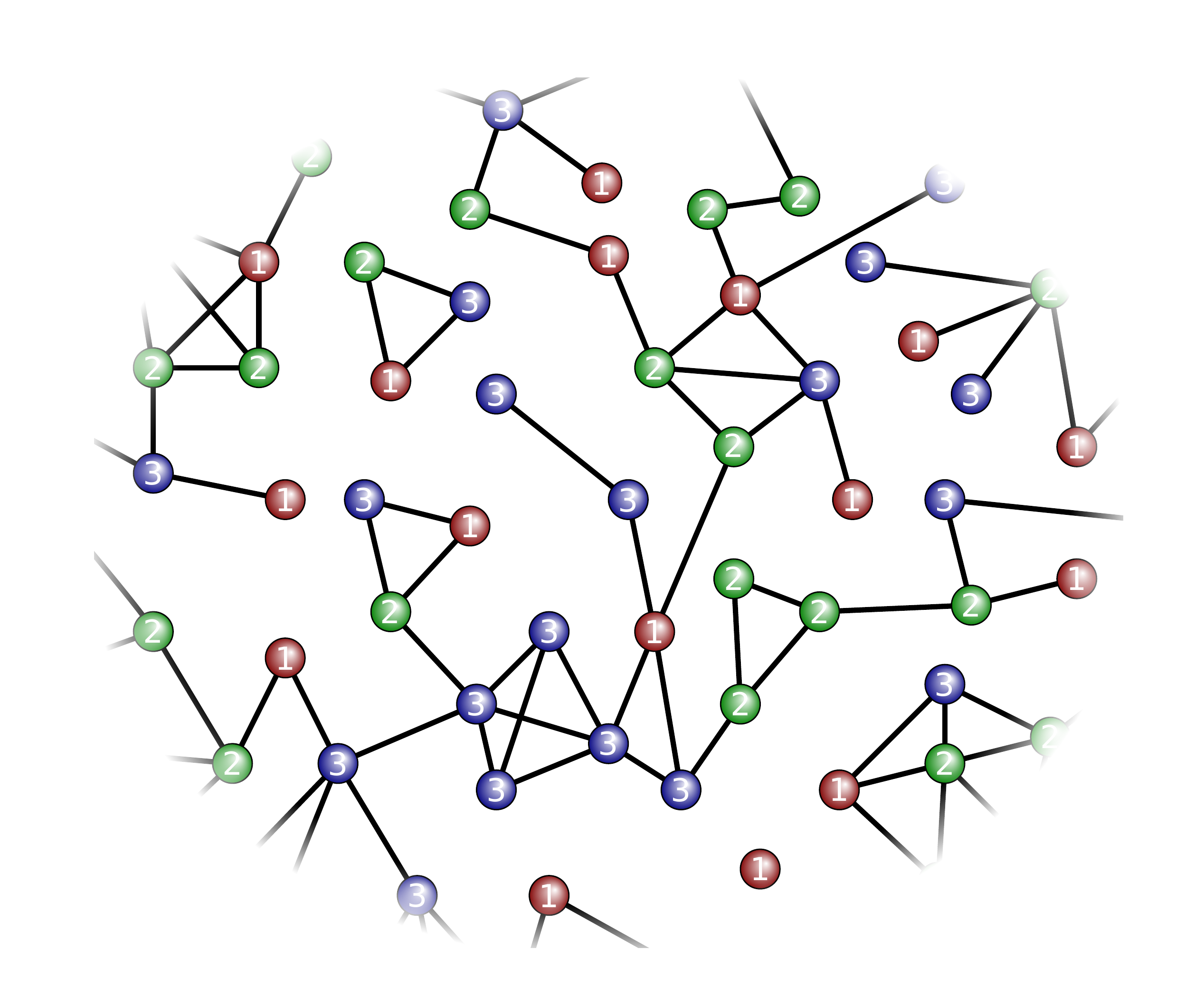}} \\
  \caption{\label{fig:omnibus_example}(Color online) Illustration of the stub matching scheme. \protect\subref{fig:omnibus_example_a} vertices are attributed a type according to the distribution $\{w_i\}$, and are given a number of stubs of each type according to the distribution $\{P_i(\bm{k})\}$. There are $N=12$ vertices, $|\mathcal{N}|=3$ types of vertices (4 vertices of type 1 in red, 4 vertices of type 2 in green, and 4 vertices of type 3 in blue), and $|\mathcal{E}|=2$ types of stubs (light blue and dark red). Stubs are then randomly matched according to a set of rules, $\mathcal{R}$, to create hyperedges. For example, a light blue stub and a dark red stub that both stem from vertices of type 2 can be matched to create a directed edge, or three light blue stubs stemming from three vertices of type 1, 2 and 3 can be matched to create a triangle. More complex hyperedges are possible and can be handled by our mathematical approach. \protect\subref{fig:omnibus_example_b} Example of a graph obtained with the stub matching scheme which can reproduce a great variety of nontrivial correlations and clustering patterns. In the infinite size limit ($N \rightarrow \infty$), the resulting graphs have an underlying tree-like structure: there are no closed path other than within clustered hyperedges.}
\end{figure*}
The CM defines an ensemble of graphs that are random in all respects except for the \textit{degree} of their vertices (the number of neighbors) which is prescribed by a given distribution $\{P(k)\}_{k\in\mathbb{N}}$. More precisely, to generate graphs of this ensemble we start with $N$ vertices and assign a degree to each one by drawing an integer from $\{P(k)\}_{k\in\mathbb{N}}$. We then build a list of stubs (half-edges) in which a vertex whose degree is $k$ appears $k$ times. We shuffle the list and pair stubs according to this randomized list to create edges. Up to corrections of order $\mathcal{O}(N^{-1})$, this procedure uniformly samples the ensemble of graphs with a given degree distribution \cite{Newman2010}. Moreover, as closed loops also occur with a probability proportional to $N^{-1}$, this procedure generates graphs that are locally tree-like in the limit $N \rightarrow \infty$.

We generalize this scheme to account for types of vertices and types of stubs. In our model, each of the $N$ vertices belongs to a type and we note $\mathcal{N}$ the set of vertex types, as in the last section. We also note $w_i$ the fraction of vertices whose type is $i$. As in the CM, vertices are assigned a number of stubs, but now these stubs are identified with types as well. We say that a vertex has $k_\alpha$ stubs of type $\alpha$, and we note $\mathcal{E}$ the set of stub types. Unless specified otherwise, Greek and Latin letters refer to types of edges and vertices, respectively. The number of stubs of each type belonging to a vertex of type $i$ is prescribed by the joint degree distribution $\{P_i(k_1,\ldots,k_{|\mathcal{E}|})\}_{k_1,\ldots,k_{|\mathcal{E}|}\in\mathbb{N}} \equiv \{P_i(\bm{k})\}_{\bm{k}\in\mathbb{N}^{|\mathcal{E}|}}$. Hence, when generating graphs from this ensemble, each of the $N$ vertices is assigned a type according to $\{w_i\}_{i\in\mathcal{N}}$ and then assigned a number of stubs of each type according to the corresponding joint degree distribution.

To generate graphs from this sequence of vertices, we build a list of stubs for each pair $(\alpha,i)$ where $\alpha\in\mathcal{E}$ and $i\in\mathcal{N}$. For example, a vertex of type $i$ that has $k_\alpha$ stubs of type $\alpha$ and $k_\beta$ stubs of type $\beta$ appears $k_\alpha$ times in the list $(\alpha,i)$ and $k_\beta$ times in the list $(\beta,i)$. Stubs are then randomly matched according to a set of rules---noted $\mathcal{R}$---to generate graphs. The information encoded in these rules is twofold. On the one hand, they prescribe from \textit{which} lists should stubs be picked during the matching step. Mathematically, this is encoded in the distribution $\{R(\bm{n})\}_{\bm{n}\in\mathbb{N}^{|\mathcal{E}|\times|\mathcal{N}|}}$ where $\bm{n}$ is a matrix whose elements, $n_{\alpha i}$ (for every $\alpha\in\mathcal{E}$ and $i\in\mathcal{N}$), give the number of stubs from each list involved in the edge (or \textit{hyperedge}, if more than two stubs are involved). The probability that an hyperedge contains $\bm{n}$ stubs is then $R(\bm{n})$.
%
% Mathematically, this is encoded in the distribution $\{R(n_{11},\ldots,n_{|\mathcal{E}|1},n_{12},\ldots,n_{|\mathcal{E}|\times|\mathcal{N}|})\}_{n_{11},\ldots,n_{|\mathcal{E}|\times|\mathcal{N}|}\in\mathbb{N}} \equiv \{R(\bm{n})\}_{\bm{n}\in\mathbb{N}^{|\mathcal{E}|\times|\mathcal{N}|}}$ giving the proportion of edges---or \textit{hyperedges}, if more than two vertices are involved---that are built by matching $n_{\alpha i}$ stubs of type $\alpha$ stemming from vertices of type $i$ (for every $\alpha\in\mathcal{E}$ and $i\in\mathcal{N}$).

On the other hand, the rules $\mathcal{R}$ prescribe \textit{how} the vertices are connected to one another within the hyperedge. For example, stubs from the list $(\alpha,i)$ and $(\alpha,j)$ could be paired to create undirected edges between layers $i$ and $j$ of \textit{multilayer} graphs. Similarly, stubs from the lists $(\beta,i)$ and $(\gamma,i)$ could be paired to create directed edges between vertices of a same type (the two types of stubs corresponding respectively to the \textit{in-}degree and \textit{out-}degree). Moreover, three stubs from a same list could be matched to create triangles, or $m$ stubs of type $\varepsilon$ stemming from different types of vertices could be matched to form a multitype Erd\H{o}s-R\'enyi motif where edges exist with probability $p$ (see Sec.~\ref{sec:omnibus_small_graphs_multitype}). In fact, the hyperedges can take any imaginable form and composition as long as they can be mapped unto the multitype random graphs defined in Sec.~\ref{sec:omnibus_small_graphs}. Note that only one stub is required to be part of an hyperedge, even if this hyperedge contributes to more than one to the degree of vertices. For instance, if stubs of type $\Delta$ correspond to triangles, a vertex with $k_\Delta=2$ will belong to two triangles. An illustration of the stub matching scheme is given in Fig.~\ref{fig:omnibus_example}.

For this graph ensemble to be consistent, the distributions $\{P_i(\bm{k})\}_{\bm{k}\in\mathbb{N}^{|\mathcal{E}|}}$ and $\{R(\bm{n})\}_{\bm{n}\in\mathbb{N}^{|\mathcal{E}|\times|\mathcal{N}|}}$ must obey certain constraints in the limit $N \rightarrow \infty$. Namely
\begin{align} \label{eq:omnibus_conditions}
  \frac{w_i\langle k_\alpha \rangle_{P_i}}{w_j\langle k_\nu \rangle_{P_j}} = \frac{\langle n_{\alpha i} \rangle_{R}}{\langle n_{\nu j} \rangle_{R}}
\end{align}
for each $i,j\in\mathcal{N}$ and $\alpha,\nu\in\mathcal{E}$, where $\langle x \rangle_Y$ represents the average of $x$ according to the distribution $Y(x)$. These constraints simply require that the ratio of the average number of elements in each list (left) equals the relative proportion in which pairs appear in hyperedges (right).

As for the CM, this \textit{stub matching} scheme uniformly samples---up to corrections of order $\mathcal{O}(N^{-1})$---a maximally random ensemble of graphs defined by the distributions $\{w_i\}_{i\in\mathcal{N}}$ and $\{P_i(\bm{k})\}_{i\in\mathcal{N};\bm{k}\in\mathbb{N}^{|\mathcal{E}|}}$, and by the rules $\mathcal{R}$. Since stubs are matched randomly, the graphs of that ensemble have an underlying tree-like structure in the limit $N \rightarrow \infty$ except within clustered hyperedges.
%
%
%
% =================================================================================================
\subsection{Probability generating functions} \label{eq:omnibus_pgf}
% =================================================================================================
%
\begin{figure*}[tb]
  \centering
  \subfloat[Before projection]{\label{fig:omnibus_pgf_a} \includegraphics[width = 0.25\linewidth]{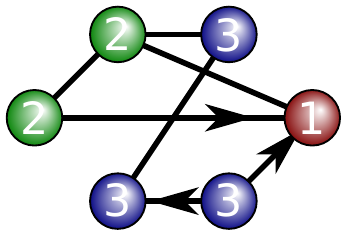}}
  \hspace{0.05\linewidth}
  \subfloat[After projection]{\label{fig:omnibus_pgf_b} \includegraphics[width = 0.25\linewidth]{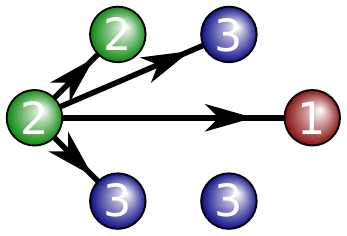}}
  \hspace{0.05\linewidth}
  \subfloat[Representation of $f_{\mu i}(\bm{x})$]{\label{fig:omnibus_pgf_c} \includegraphics[width = 0.25\linewidth]{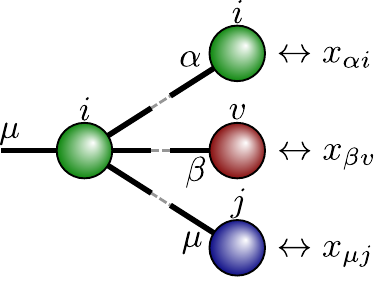}} \\
  \caption{\label{fig:omnibus_pgf}(Color online) \protect\subref{fig:omnibus_pgf_a}--\protect\subref{fig:omnibus_pgf_b} Effective tree-like structure of an hyperedge from the point of view of a vertex of type 2. There exists an effective edge between the initial vertex of type 2 and any vertices that are directly or indirectly reachable from it. The probability for an effective edge to exist corresponds to the probability that a direct or indirect path exists. \protect\subref{fig:omnibus_pgf_c} Schematic representation of the pgf $f_{\mu i}(\bm{x})$. Knowing that a vertex of type $i$ has been reached from one of its stubs of type $\mu$ [i.e., a pair $(\mu,i)$], this pgf generates the distribution of the number of vertices of each type in its neighborhood, as well as the type of stubs from which they have been reached. The types of the vertices and of the stubs are identified with the subscripts of the variables $\bm{x}=\{x_{\nu i}\}$.}
\end{figure*}
To solve percolation on this general ensemble of random graphs, we adapt the well-known pgf approach \cite{Allard12_JPhysA,Newman01_PhysRevE} to account for vertex and stub types. As mentioned above, this approach assumes that the structure of the graphs is locally tree-like, an assumption that is not valid whenever an hyperedge contains a loop (e.g., a triangle). However, by solving the component size distribution on each hyperedge beforehand, it is possible to consider that the hyperedge has an \textit{effective} tree-like structure: the probability that there is an \textit{effective} edge from vertex A to vertex B is simply the probability that vertex B can be reached from vertex A either directly or through the other vertices in the hyperedge. Figures~\ref{fig:omnibus_pgf}\protect\subref{fig:omnibus_pgf_a}--\protect\subref{fig:omnibus_pgf_b} illustrate the idea behind the effective tree-like structure. This slight change of perspective allows the use of the pgf approach even though the tree-like structure assumption is not valid in the original graph ensemble.

The effective tree-like structure of hyperedges is unveiled with Eqs.~\eqref{eq:omnibus_gnp}, where a vertex is now identified by the pair $(\alpha,i)$ instead of by its vertex type solely. In other words, we keep track of the type of the vertices but also the type of the stubs through which they are involved in the hyperedge. As a result, bold variables like $\bm{n}$ and $\bm{l}$ now contain $|\mathcal{E}|\times|\mathcal{N}|$ elements instead of the $|\mathcal{N}|$ elements as in Sec.~\ref{sec:omnibus_small_graphs}. The quantity $Q_{\alpha i}(\bm{l}|\bm{n};\mathcal{R})$ therefore corresponds to the probability that a pair $(\alpha,i)$ leads to $\bm{l}$ pairs---i.e., $l_{\nu j}$ pairs $(\nu,j)$, for each $\nu\in\mathcal{E}$ and $j\in\mathcal{N}$---in an hyperedge containing $\bm{n}$ pairs. A dependency on the rules $\mathcal{R}$ has been added in $Q_{\alpha i}(\bm{l}|\bm{n};\mathcal{R})$ to explicitly mark that the inner structure of the hyperedges (e.g., quenched or random nature, probabilities of existence of vertices or edges) is prescribed by these rules \footnote{It is not forbidden for two hyperedges to have the same composition $\bm{n}$ while having different nature or different rules of connection. In such case, the probabilities $Q_{\alpha i}(\bm{l}|\bm{n};\mathcal{R})$ should represent the appropriate weighted average of the probabilities computed with Eqs.~\eqref{eq:omnibus_gnp} for each hyperedge. It is however strongly encouraged to use different types of stubs---hence different compositions---for each different categories of hyperedges to keep the analysis as clear as possible.}.

The pgf that generates the distribution of the number of pairs that can be reached from an initial pair $(\alpha,i)$ in an hyperedge containing $\bm{n}$ pairs is
\begin{align} \label{eq:omnibus_theta_tmp}
  \sum_{\bm{l}=\bm{\delta}_{\alpha} \circ \bm{\delta_i}}^{\bm{n}} Q_{\alpha i}(\bm{l}|\bm{n};\mathcal{R}) \prod_{\substack{\nu \in \mathcal{E}\\j\in\mathcal{N}}}x_{\nu j}^{l_{\nu j}-\delta_{\alpha\nu}\delta_{ij}}
\end{align}
where the sum covers every possible instances of $\bm{l}$ such that $\delta_{\mu\alpha}\delta_{mi} \leq l_{\mu m} \leq n_{\mu m}$ for every $m \in \mathcal{N}$ and $\mu \in \mathcal{E}$ (``$\circ$'' denotes the entrywise product). These two deltas of Kronecker account for the fact that there is at least one pair $(\alpha,i)$ in the hyperedge. Similarly, the two others deltas of Kronecker $\delta_{\alpha\nu}\delta_{ij}$ appearing in Eq.~\eqref{eq:omnibus_theta_tmp} remove the initial pair---by definition included in $\bm{l}$---from the count of reachable pairs. Because we are ultimately interested in the number of pairs that can be reached from a given initial pair regardless of the specifics of the hyperedge, we must remove the dependency of Eqs.~\eqref{eq:omnibus_theta_tmp} on the composition $\bm{n}$. To do so, we average this pgf over the probabilities that the initial pair $(\alpha,i)$ belongs to an hyperedge whose composition is $\bm{n}$. Since stubs are matched randomly, a vertex identified by the pair $(\alpha,i)$ is ten times more likely to belong to an hyperedge containing ten pairs $(\alpha,i)$ than to belong to an hyperedge that contains only one pair $(\alpha,i)$. Consequently the probabilities $R(\bm{n})$ must be weighted by the number of pairs $(\alpha,i)$ that each composition contains, i.e., averaged over $n_{\alpha i}R(\bm{n})/\langle n_{\alpha i} \rangle_{R}$, where the normalizing factor, $\langle n_{\alpha i} \rangle_{R}$, is the average value of $n_{\alpha i}$ with respect to the distribution $R(\bm{n})$. Doing so yields the pgf generating the distribution of the number of pairs of each types that can be reached from a pair $(\alpha,i)$
\begin{align} \label{eq:omnibus_theta}
  \theta_{\alpha i}(\bm{x}) & = \sum_{\bm{n}} \frac{n_{\alpha i} R(\bm{n})}{\langle n_{\alpha i} \rangle_{R}} \sum_{\bm{l}=\bm{\delta_\alpha} \circ \bm{\delta_i}}^{\bm{n}} Q_{\alpha i}(\bm{l}|\bm{n};\mathcal{R}) \prod_{\substack{\nu \in \mathcal{E}\\j\in\mathcal{N}}}x_{\nu j}^{l_{\nu j}-\delta_{\alpha\nu}\delta_{ij}} \ ,
\end{align}
where the sum over $\bm{n}$ covers all hyperedge compositions such that $R(\bm{n}) \neq 0$. Computed for each initial pair $(\alpha,i)$, $\theta_{\alpha i}(\bm{x})$ provides the projection of the outcomes of percolation on the hyperedges unto an effective branching tree and therefore permits the use of the pgf approach.

To solve percolation on the graphs defined in the previous subsection, we first need to compute the distribution of the composition of the neighborhood of vertices. The neighborhood of a vertex is the set of \textit{reachable} vertices with which it shares an hyperedge. In other words, vertex B is a neighbor of vertex A if there exists an effective edge from vertex A to vertex B. The pgf $\theta_{\alpha i}(\bm{x})$ generates the distribution of neighbors that a vertex of type $i$ has through one of its stubs of type $\alpha$. In the limit $N \rightarrow \infty$, the tree-like structure of the graphs ensures that the neighboring vertices reachable through two different stubs do not overlap. Hence, the composition of the neighborhood of a vertex of type $i$ that has $k_\alpha$ and $k_\beta$ stubs of type $\alpha$ and $\beta$ is generated by $\big[\theta_{\alpha i}(\bm{x})\big]^{k_\alpha} \cdot \big[\theta_{\beta i}(\bm{x})\big]^{k_\beta}$, which corresponds to the convolution of the distributions. Since the number of stubs belonging to vertices of type $i$ is distributed according to $\{P_i(\bm{k})\}_{\bm{k}\in\mathbb{N}^{|\mathcal{E}|}}$, we obtain that the distribution of the composition of the neighborhood of vertices of type $i$ is generated by
\begin{align} \label{eq:omnibus_gi}
  g_i(\bm{x}) & = \sum_{\bm{k}} P_i(\bm{k}) \prod_{\alpha\in\mathcal{E}} \big[  \theta_{\alpha i}(\bm{x}) \big]^{k_\alpha} \ ,
\end{align}
where the sum covers all cases where $P_i(\bm{k}) \neq 0$. As in $\theta_{\alpha i}(\bm{x})$, this pgf keeps track of the type of the stubs from which the neighboring vertices have been reached through the subscripts of the variables $\bm{x} = \{x_{\nu j}\}_{\nu\in\mathcal{E};j\in\mathcal{N}}$. In other words, $g_i(\bm{x})$ generates the number of pairs that are in the neighborhood of a vertex of type $i$. This pgf is analogous to the function $G_0(x)$ generating the degree distribution in the CM \cite{Newman2010,Newman01_PhysRevE}.

The complete solution to the percolation problem requires the distribution of possible neighborhood compositions for vertices reached through one of their stubs. As discussed for $\theta_{\alpha i}(\bm{x})$, the probability for a stub of type $\mu$ to be attached to a vertex with a total of $\bm{k}$ stubs (i.e., $k_\alpha$ stubs of type $\alpha$ for every $\alpha\in\mathcal{E}$) is weighted by the number of stubs of type $\mu$ that this vertex has. Hence, given that a vertex of type $i$ has been reached through one of its stubs of type $\mu$ [i.e., a pair $(\mu,i)$], the composition of the neighborhood accessible from its \textit{other} stubs is generated by
\begin{align} \label{eq:omnibus_fmui}
 f_{\mu i}(\bm{x}) & = \sum_{\bm{k}} \frac{k_{\mu}P_i(\bm{k})}{\langle k_\mu \rangle_{P_i}} \prod_{\alpha\in\mathcal{E}} \big[  \theta_{\alpha i}(\bm{x}) \big]^{k_\alpha - \delta_{\alpha\mu}} \ ,
\end{align}
where the delta $\delta_{\alpha\mu}$ has been added to exclude from the count the stub of type $\mu$ from which the vertex has been reached, and where $\langle k_\mu \rangle_{P_i}$ is the average number of stubs of type $\mu$ that vertices of type $i$ have. The distributions generated by $f_{\mu i}(\bm{x})$ are analogous to the \textit{excess} degree distribution generated by $G_1(x)$ in the CM \cite{Newman2010,Newman01_PhysRevE}. Figure~\ref{fig:omnibus_pgf}\protect\subref{fig:omnibus_pgf_c} illustrates the information encoded in the pgfs $f_{\mu i}(\bm{x})$.
%
%
%
% =================================================================================================
\subsection{Extensive ``giant'' component} \label{sec:omnibus_extensive_component}
% =================================================================================================
%
Having defined the pgfs $g_i(\bm{x})$ and $f_{\mu i}(\bm{x})$, the behavior of the extensive ``giant'' component can be predicted in the limit $N \rightarrow \infty$ using simple self-consistency arguments. We define $a_{\mu i}$ as the probability that a vertex of type $i$ reached via one of its stubs of type $\mu$ does \textit{not} lead to the giant component. Self-consistency then requires that if this pair does not lead to the giant component, then neither should the pairs that are reachable from it. Since the distribution of the number of pairs reachable from a given pair $(\mu,i)$ is generated by Eq.~\eqref{eq:omnibus_fmui}, this self-consistency requirement can be rewritten as
\begin{align} \label{eq:omnibus_amui}
  a_{\mu i} & = f_{\mu i}(\bm{a})
\end{align}
for every $\mu\in\mathcal{E}$ and $i\in\mathcal{N}$. Because the coefficients of $f_{\mu i}(\bm{x})$ are normalized (they form a probability distribution), the point $\bm{a}=\bm{1}$ (every $a_{\mu i}$ equals 1) is always a solution of Eqs.~\eqref{eq:omnibus_amui}. However, as the density of edges and/or vertices increases with increasing $\{r_j\}_{j\in\mathcal{N}}$ and/or $\{p_{jk}\}_{j,k\in\mathcal{N}}$, another solution where at least one element of $\bm{a}$ is smaller than 1 appears. This new solution marks the emergence of an extensive component.

Because their coefficients are all positives, the pgfs $f_{\mu i}(\bm{x})$ are all convex and monotonic increasing in $[0,1]^{|\mathcal{E}|\times|\mathcal{N}|}$. Hence when $\bm{a}=\bm{1}$ is the only solution of Eqs.~\eqref{eq:omnibus_amui} in $[0,1]^{|\mathcal{E}|\times|\mathcal{N}|}$, it is the stable fixed point of (with $n\in\mathbb{N}$)
\begin{align} \label{eq:omnibus_amap}
  \bm{a}^{(n+1)} & = \bm{f}\big(\bm{a}^{(n)}\big) \ ,
\end{align}
for any initial condition $\bm{a}^{(0)}$ in $[0,1]^{|\mathcal{E}|\times|\mathcal{N}|}$, and where the map $\bm{f}(\bm{x})$ consists of every $f_{\mu i}(\bm{x})$. This fixed point becomes unstable through a transcritical bifurcation as soon as another solution in $[0,1]^{|\mathcal{E}|\times|\mathcal{N}|}$ appears. The shape of $\bm{f}(\bm{x})$ in $[0,1]^{|\mathcal{E}|\times|\mathcal{N}|}$ and the fact that $\bm{f}(\bm{1})=\bm{1}$ implies that this other solution is unique in the interval of interest, that it is a stable fixed point of Eq.~\eqref{eq:omnibus_amap}, and that the transition is continuous. Analyzing the stability of $\bm{f}(\bm{x})$ around the fixed point $\bm{a}=\bm{1}$ leads to the criterion for the emergence of the giant component
\begin{align} \label{eq:omnibus_seuil}
  \mathrm{det}(\mathbf{J}-\mathbf{I}) = 0 \ ,
\end{align}
where $\mathbf{J}$ is the Jacobian matrix of $\bm{f}(\bm{x})$ around $\bm{x}=\bm{1}$, and $\mathbf{I}$ is the identity matrix. Put differently, an extensive component exists whenever the largest eigenvalue of $\mathbf{J}$, $\lambda_\mathrm{max}(\mathbf{J})$, is greater than one \footnote{Because the coefficients of $\bm{f}(\bm{x})$ are all positives, $\mathbf{J}$ is a non-negative matrix and the Perron-Frobenius theorem ensures that its largest eigenvalue is real, positive and non-degenerate.}.

Having solved Eqs.~\eqref{eq:omnibus_amui}, the probability that a vertex of type $i$ leads to the giant component through at least one of its neighbors is given by $\mathcal{P}_i=1-g_i(\bm{a})$. Consequently, the probability that a randomly chosen vertex does lead to the giant component is
\begin{align} \label{eq:omnibus_P}
 \mathcal{P} & = \sum_{i\in\mathcal{N}} \frac{r_iw_i \mathcal{P}_i}{\sum_{j\in\mathcal{N}}r_jw_j} = 1 - \sum_{i\in\mathcal{N}} \frac{r_iw_ig_i(\bm{a})}{\sum_{j\in\mathcal{N}}r_jw_j}  \ ,
\end{align}
where $r_i$ is the probability that a vertex of type $i$ exists.

As shown in Sec.~\ref{sec:omnibus_small_graphs}, hyperedges may include directed edges, or edges that are more likely to exist in one direction than the other [i.e., $p_{ij} \neq p_{ji}$ in Eqs.~\eqref{eq:omnibus_gnp_1}]. This implies that while vertex B is in the neighborhood of vertex A, vertex A may not be in the neighborhood of vertex B. From such local asymmetries, a global asymmetry arises between the probability that a vertex leads to the giant component, $\mathcal{P}$, and the relative size $\mathcal{S}$ of the giant component. In such case, the extensive component has a ``bow-tie'' structure \cite{Newman01_PhysRevE,Allard09_PhysRevE} meaning that the vertices involved in the extensive component belong to one of the three non-overlapping sets $\mathcal{I}^{\mathrm{in}}$, $\mathcal{I}^{\mathrm{both}}$ and $\mathcal{I}^{\mathrm{out}}$. The set $\mathcal{I}^{\mathrm{in}}$ includes vertices that lead to the giant component but that cannot be reached from it; these vertices are somehow ``hidden'' behind directed edges. The set $\mathcal{I}^{\mathrm{out}}$ contains vertices that cannot lead to the giant component but that can be reached from it; they are positioned downstream of directed edges. The set $\mathcal{I}^{\mathrm{both}}$ contains vertices that lead to the giant component and that can be reached from it. From this, we conclude $\mathcal{P}=|\mathcal{I}^{\mathrm{in}}\bigcup\mathcal{I}^{\mathrm{both}}|/N$ and $\mathcal{S}=|\mathcal{I}^{\mathrm{both}}\bigcup\mathcal{I}^{\mathrm{out}}|/N$.

This perspective offers a direct and intuitive way to calculate $\mathcal{S}$: it is the probability that a vertex does not lead to the extensive component when the direction of every edges is reversed. This edge reversal is fully encoded in $\bar{Q}_{\alpha i}(\bm{l}|\bm{n};\mathcal{R})$ computed with Eqs.~\eqref{eq:omnibus_gnp} with incoming directed edges swapped into outgoing ones (and vice versa), and with edges that were more likely to exist in a given direction now more likely to exist in the opposite direction (i.e., $p_{ij}$ becomes $p_{ji}$). From these probabilities, we define the pgfs $\bar{\theta}_{\alpha i}(\bm{x})$, $\bar{g}_i(\bm{x})$ and $\bar{f}_{\mu i}(\bm{x})$ which are analogous to the ones previously defined [$P_i(\bm{k})$ and $R(\bm{n})$ remain unchanged]. Defining $\bar{a}_{\mu i}$ as the probability that a vertex of type $i$ reached by one of its stubs of type $\mu$ does not lead to the giant component in the \textit{reversed} graph ensemble, self-consistency now requires
\begin{align} \label{eq:omnibus_bara_mui}
  \bar{a}_{\mu i} & = \bar{f}_{\mu i}(\bm{\bar{a}})
\end{align}
for every $\mu\in\mathcal{E}$ and $i\in\mathcal{N}$. As for Eqs.~\eqref{eq:omnibus_amui}, the solution of this set of equations correspond to the fixed point of the corresponding map and can therefore be obtained by successive iterations of any initial condition in $[0,1]^{|\mathcal{E}|\times|\mathcal{N}|}$. The elements of the Jacobian matrix of both Eqs.~\eqref{eq:omnibus_amui} and \eqref{eq:omnibus_bara_mui} are the average number of pairs, say $(\alpha,j)$, that are in the neighborhood of a pair, say $(\mu,i)$, in their respective graph ensemble. Since both systems, Eqs.~\eqref{eq:omnibus_amui} and \eqref{eq:omnibus_bara_mui}, correspond to different perspectives of the same graph ensemble, the two Jacobian matrices are linked by a similarity transformation, and therefore have the same eigenvalues. Hence the transcritical bifurcation occurs simultaneously in both systems.

Having obtained $\bm{\bar{a}}$ from Eqs.~\eqref{eq:omnibus_bara_mui}, the probability for a vertex of type $i$ to be part of the giant component is $\mathcal{S}_i=1-\bar{g}_i(\bm{\bar{a}})$, and the relative size of the giant component is
\begin{align} \label{eq:omnibus_S}
  \mathcal{S} = \sum_{i\in\mathcal{N}} \frac{r_iw_i \mathcal{S}_i}{\sum_{j\in\mathcal{N}}r_jw_j} = 1 - \sum_{i\in\mathcal{N}} \frac{r_iw_i\bar{g}_i(\bm{\bar{a}})}{\sum_{j\in\mathcal{N}}r_jw_j} \ .
\end{align}
Clearly, when all hyperedges are symmetric (i.e., $p_{ij}=p_{ji}$ for every $i,j\in\mathcal{N}$) there is no global asymmetry in the graph ensemble, and $\mathcal{P}=\mathcal{S}$. Also, whenever Eqs.~\eqref{eq:omnibus_theta_tmp}--\eqref{eq:omnibus_S} are used in the context of site percolation---where vertices exist or are activated with a given set of probabilities---the value of $\mathcal{P}$ and $\mathcal{S}$ is relative to the number of vertices that exist. In other words, $\mathcal{P}$ is the probability that an existing vertex leads to an extensive component, and $\mathcal{S}$ is the probability that an existing vertex is part of it. 
%
%
%
% =================================================================================================
\subsection{Small components}
% =================================================================================================
%
Substituting $x_{\nu j}$ by $z_j$ for every $j\in\mathcal{N}$ and $\nu\in\mathcal{E}$ in Eq.~\eqref{eq:omnibus_gi} yields a pgf that generates the number of vertices of each type that are \textit{directly accessible} from a vertex of type $i$ (i.e., vertices that are in its neighborhood). In other words, the information concerning the types of stubs is lost. Using self-consistency arguments similar to the one used in the previous subsection, it is possible to obtain a pgf that generates the distribution of the number of vertices of each type that will be \textit{eventually reached} from a vertex of type $i$; the \textit{reach} of this new pgf is no longer limited to the immediate neighborhood. In fact this new pgf allows to investigate the composition and the sizes of the components that contain a finite number of vertices. Let this new pgf be denoted $K(\bm{z})$.

To compute $K(\bm{z})$, we first consider the pgf $A_{\alpha i}(\bm{x})$ that generates the distribution of the number of all pairs of each type that will \textit{eventually} be reached (i.e., not limited to the first neighbors) from a vertex of type $i$ given that this vertex has been reached from one of its stubs of type $\alpha$. In other words, this function generates the distribution of the number of the vertices that are eventually reached from a pair $(\alpha,i)$. Note that $A_{\alpha i}(\bm{x})$ is a function of $\bm{x}$ so that it keeps track of the type of the stubs from which each vertex has been reached. Besides yielding a tree-like structure, the \textit{stub matching} scheme used to generate graphs implies that the pgfs $\{f_{\mu i}(\bm{x})\}$ are invariant under translations on the graphs in the limit $N \rightarrow \infty$. In other words, while navigating on a graph from this ensemble, the number and the type of the vertices downstream from any given vertex does not depend on the types of the vertices (or the types of the stubs) previously encountered; navigating on graphs from this ensemble is a stationary Markov process (i.e., it only depends on the \textit{current position} on the graph). Consequently, a vertex of type $i$ reached from one of its stubs of type $\alpha$ and a pair $(\alpha,i)$ present in its neighborhood should both lead to a finite tree whose size and composition are identically distributed; this distribution is generated by $A_{\alpha i}(\bm{x})$. Considering every combination $(\alpha,i)$, this self-consistency requirement can be mathematically formulated as
\begin{align} \label{eq:omnibus_Amui}
  A_{\alpha i}(\bm{x}) & = x_{\alpha i} f_{\alpha i} \big( \bm{A}(\bm{x}) \big) \ ,
\end{align}
where the extra $x_{\alpha i}$ accounts for the vertex of type $i$ that has been reached through one of its stubs of type $\alpha$. Analogously to the set of probabilities $\{a_{\alpha i}\}$, the pgfs $\{A_{\alpha i}(\bm{x})\}$ are the fixed point of (with $n\in\mathbb{N}$)
\begin{align} \label{eq:omnibus_Amap}
  \bm{A}^{(n+1)}(\bm{x}) & = \bm{x} \circ \bm{f} \big( \bm{A}^{(n)}(\bm{x}) \big) \, 
\end{align}
where ``$\circ$'' denotes the entrywise product, and where $\bm{f(x)}$ is the same map as in Eq.~\eqref{eq:omnibus_amap}. It is in fact straightforward to show that the extra $\bm{x}$ guarantees that the distributions generated by $\bm{A}(\bm{x})$ can be obtained for components of $n$ vertices or less in $n+1$ iterations of Eq.~\eqref{eq:omnibus_Amap} from the initial condition $\bm{A}^{(0)}(\bm{x})=\bm{1}$ [i.e., $A_{\alpha i}^{(0)}(\bm{x})=1$ for every $i\in\mathcal{N}$ and $\alpha\in\mathcal{E}$].

Having obtained $\bm{A}(\bm{x})$ up to a sufficient size of components, $n$, the number of vertices of each type that can be reached in a finite component from a randomly chosen vertex of type $i$ is generated by $K_i(\bm{z}) \equiv z_i g_i \big( \bm{A(z)} \big)$. The pgf generating the number of vertices of each type that are accessible in a small component from a randomly chosen (existing) vertex is
\begin{align} \label{eq:omnibus_K}
  K(\bm{z}) & = \sum_{i\in\mathcal{N}} \frac{r_iw_iK_i(\bm{z})}{\sum_{j\in\mathcal{N}}r_jw_j} = \sum_{i\in\mathcal{N}} \frac{r_iw_iz_i g_i \big( \bm{A}(\bm{z})\big)}{\sum_{j\in\mathcal{N}}r_jw_j} \ .
\end{align}
It is worth mentioning that the distributions generated by $K(\bm{z})$ and $\{A_{\alpha i}(\bm{z})\}$ are not normalized in the presence of an extensive component as there is a non-zero probability that a pair $(\alpha,i)$ leads to the giant component. In fact, comparing Eqs.~\eqref{eq:omnibus_amap} and \eqref{eq:omnibus_Amap} leads to the conclusion that $A_{\alpha i}(\bm{1}) = a_{\alpha i}$ and that $K(\bm{1})=1-\mathcal{P}$.
%
%
%
%
%
% =================================================================================================
\section{Special cases and applications} \label{sec:omnibus_applications}
% =================================================================================================
%
To demonstrate the versatility and the flexibility of the formalism, we present a series of representative examples. This will also clarify the conceptual and numerical steps necessary to implement such a general approach.
%
%
%
% =================================================================================================
\subsection{Semi-directed random graphs} \label{sec:omnibus_example_semidirected}
% =================================================================================================
%
Semi-directed random graphs are composed of indistinguishable vertices connected via undirected and directed edges. They were used in Ref.~\cite{Meyers06_JTheorBiol} to study the impact of non-reciprocal connections in contact networks on the propagation of an emerging infectious disease. These non-reciprocal connections accounted for the susceptibility of health-care workers to get infected from infectious individuals seeking treatments in hospitals. Semi-directed are also a good first example for they have the well-known undirected graphs and directed graphs as special cases.

Every vertices in these graphs belong to the same type ($|\mathcal{N}|=1$, type 1, $w_1$=1), and there are $|\mathcal{E}|=3$ types of stubs: stubs of type A are paired together to form undirected edges, and stubs of type B (outgoing) and C (incoming) are paired to form a directed edge. The joint degree distribution $P_1(\bm{k})=P_1(k_A,k_B,k_C)$ corresponds to the distribution of \textit{undirected} degree, \textit{out}-degree and \textit{in}-degree. In this scenario, the conditions given by Eq.~\eqref{eq:omnibus_conditions} imply that there must be as much incoming stubs as there are outgoing stubs, $\langle k_B \rangle_{P_1} = \langle k_C \rangle_{P_1}$, and they fix the values of $R(\bm{n})=R(n_{A1},n_{B1},n_{C1})$ in terms of the average degrees, $R(2,0,0)=1-R(0,1,1)=\langle k_A \rangle_{P_1} / (\langle k_A \rangle_{P_1}+2\langle k_B \rangle_{P_1})$. Assuming that edges exist with probability $p_{11}$ and vertices exist with probability $r_1$, we find from Eqs.~\eqref{eq:omnibus_gnp} and \eqref{eq:omnibus_theta}
\begin{subequations} \label{eq:omnibus_theta_semidirected}
\begin{align}
  \theta_{A1}(\bm{x}) & = (1-r_1p_{11}) + r_1p_{11}x_{A1} \\
  \theta_{B1}(\bm{x}) & = (1-r_1p_{11}) + r_1p_{11}x_{C1} \\
  \theta_{C1}(\bm{x}) & = 1 \ ,
\end{align}
\end{subequations}
from which we define the pgfs $g_1(\bm{x})$, $f_{A1}(\bm{x})$ and $f_{C1}(\bm{x})$ from Eqs.~\eqref{eq:omnibus_gi} and \eqref{eq:omnibus_fmui}. Note that $f_{B1}(\bm{x})$ does not exist as vertices cannot be reached by an outgoing stub. Similarly, when reversing the direction of edges (directed edges now run from C stubs to B stubs), we obtain
\begin{subequations} \label{eq:omnibus_bartheta_semidirected}
\begin{align}
  \bar{\theta}_{A1}(\bm{x}) & = (1-r_1p_{11}) + r_1p_{11}x_{A1} \\
  \bar{\theta}_{B1}(\bm{x}) & = 1                 \\
  \bar{\theta}_{C1}(\bm{x}) & = (1-r_1p_{11}) + r_1p_{11}x_{B1} \ ,
\end{align}
\end{subequations}
which yield the pgfs $\bar{g}_1(\bm{x})$, $\bar{f}_{A1}(\bm{x})$ and $\bar{f}_{B1}(\bm{x})$ [$\bar{f}_{C1}(\bm{x})$ is non-defined]. Using Eqs.~\eqref{eq:omnibus_theta_semidirected} and \eqref{eq:omnibus_bartheta_semidirected} in Eqs.~\eqref{eq:omnibus_gi}--\eqref{eq:omnibus_K} with $r_1=1$ yields the results obtained in Ref.~\cite{Meyers06_JTheorBiol}, and to the ones obtained for purely directed \cite{Newman01_PhysRevE} or purely undirected random graphs \cite{Newman02_PhysRevE} in the appropriate limits.
%
%
%
% =================================================================================================
\subsection{Correlated random graphs} \label{sec:omnibus_example_correlated}
% =================================================================================================
%
Other interesting special cases of our model are correlated random graphs: graphs where vertices are more likely to be connected with vertices having specific intrinsic properties (e.g., degree, centrality, ethnicity, age group, gender). In such cases, there are $|\mathcal{N}|$ types of vertices, one for each intrinsic property, and there are as many types of stubs: each type of stubs corresponds to the type of the vertex that is at the other end of the edge. To simplify the notation, types of stubs will be identified by the type of the vertex toward which they point (i.e., $\mathcal{E}=\mathcal{N}$). Hence the joint degree distribution $P_i(\bm{k})$ prescribes the number of vertices of each type that vertices of type $i$ are connected to. The conditions \eqref{eq:omnibus_conditions} ask that there are as many stubs stemming from vertices of type $i$ toward vertices of type $j$ as in the reverse direction, $w_i\langle k_j \rangle_{P_i}=w_j\langle k_i \rangle_{P_j}$. These constraints also prescribe the distribution
\begin{align}
  R(\bm{n})= \frac{1}{R'} \sum_{i,j\in\mathcal{N}} (1-\delta_{0,n_{ji}}) w_i \langle k_j \rangle_{P_i} \ ,
\end{align}
where $R'=\sum_{i',j'\in\mathcal{N}} w_{i'} \langle k_{j'} \rangle_{P_{j'}}$ is simply the normalization factor. Assuming that vertices of type $i$ exist with probability $r_i$, and that edges going from a vertex of type $i$ to a vertex of type $j$ exist with probability $p_{ij}$ (i.e., edges may be more likely to exist in one direction that in the other), we get from Eqs.~\eqref{eq:omnibus_gnp} and \eqref{eq:omnibus_theta}
\begin{subequations} \label{eq:omnibus_theta_correlated}
\begin{align}
  \theta_{ji}(\bm{x}) & = (1 - r_j p_{ij}) + r_j p_{ij} x_{ij} \label{eq:omnibus_theta_correlated_a} \\
  \bar{\theta}_{ji}(\bm{x}) & = (1 - r_j p_{ji}) + r_j p_{ji} x_{ij} \label{eq:omnibus_theta_correlated_bara}
\end{align}
\end{subequations}
for $i,j\in\mathcal{N}$. Using Eqs.~\eqref{eq:omnibus_theta_correlated} in Eqs.~\eqref{eq:omnibus_gi}--\eqref{eq:omnibus_K} and setting every $r_i = 1$ yields the results obtained in Ref.~\cite{Allard09_PhysRevE} for \textit{multitype} graphs, which are themselves a generalization of several other formalisms \cite{Newman02_PhysRevE,Newman03a_PhysRevE,Newman01_PhysRevE}. We have also used this approach in Ref.~\cite{Allard2014} to study the observability of random graphs, and in Ref.~\cite{Hebert-Dufresne13_PhysRevE} to define an ensemble of graphs with an arbitrary k-core structure.
%
%
%
% =================================================================================================
\subsection{Degree-correlated random graphs} \label{sec:omnibus_example_degree-correlated}
% =================================================================================================
%
An important category of correlations is the one based on the degree of vertices \cite{Newman02_PhysRevLett,Vazquez03_PhysRevE}. These correlations are encoded in the conditional probability $P(d'|d)$ corresponding to the probability that the neighbor of a vertex with a degree $d$ has a degree equal to $d'$. This can be reproduced with our formalism by considering that every vertex with the same degree are of the same type (i.e., a vertex of type $i$ has $i$ neighbors, and consequently $\{w_i\}$ corresponds to the degree distribution), and by using the following type-specific joint degree distribution
\begin{align} \label{eq:omnibus_Pi_CCM}
  P_i(\bm{k}) & = \frac{i!}{\prod_{j'\in\mathcal{N}}k_{j'}!} \prod_{j\in\mathcal{N}}[P(j|i)]^{k_j} \ .
\end{align}
From Eq.~\eqref{eq:omnibus_gi}, we obtain
\begin{align} \label{eq:omnibus_gi_CCM}
  g_i(\bm{x}) & = \left[ \sum_{j\in\mathcal{N}} P(j|i) \theta_{ji}(\bm{x}) \right]^i \ ,
\end{align}
where $\theta_{ji}(\bm{x})$ is given by Eq.~\eqref{eq:omnibus_theta_correlated_a}, and Eq.~\eqref{eq:omnibus_fmui} yields
\begin{align} \label{eq:omnibus_fmui_CCM}
  f_{li} (\bm{x}) & = \left[ \sum_{j\in\mathcal{N}} P(j|i) \theta_{ji}(\bm{x}) \right]^{i-1} \ ,
\end{align}
which is independent of the type of the vertex/stub, namely $l$, from which the vertex has been reached. This is a direct consequence of the multinomial distribution in Eq.~\eqref{eq:omnibus_Pi_CCM} and shows that our approach, through the joint distribution $P_i(\bm{k})$, can include more detailed correlations in the degree of the neighbors of vertices.

It may be useful at this point to illustrate the precise connection with previous works. Consider the quantity $u_i\equiv\sum_{j\in\mathcal{N}}P(j|i)[(1-r_jp_{ji})+r_jp_{ji}\bar{a}_{ij}]$ which under successive application of Eqs.~\eqref{eq:omnibus_bara_mui}, \eqref{eq:omnibus_theta_correlated_bara} and \eqref{eq:omnibus_fmui_CCM} becomes the self-consistent expression
\begin{align} \label{eq:omnibus_u_i}
  u_i = \sum_{j \in \mathcal{N}} P(j|i) \big[ (1-r_jp_{ji}) + r_j p_{ji} u_j^{j-1} \big]
\end{align}
for every $i \in \mathcal{N}$. Setting every $r_jp_{ji}=1-f$, with $0 \leq f \leq 1$, in this last equation yields Eqs.~(5) and (13) of Ref.\cite{Vazquez03_PhysRevE}, while Eq.~(8) of Ref.~\cite{Newman02_PhysRevLett} is obtained by setting $r_jp_{ji}=1$. Similarly, replacing $P(j|i)$ by $jw_j/\sum_{l\in\mathcal{N}}lw_l$ and assuming every $u_i=u$ in Eq.~\eqref{eq:omnibus_u_i} yields the results of Ref.~\cite{Callaway00_PhysRevLett}. More precisely, setting every $p_{ji}=1$ allows to retrieve their Eq.~(15), and their Eq.~(8) is obtained by setting every $r_jp_{ji}=q_sq_b$, with $0 \leq q_s,q_b \leq 1$. Expressions for the size of the extensive component derived in Refs.~\cite{Callaway00_PhysRevLett,Newman02_PhysRevLett,Vazquez03_PhysRevE} can be obtained from our formalism similarly.
% Setting every $r_i=r$ and every $p_{ij}=p$ in Eqs.~\eqref{eq:omnibus_theta_correlated}--\eqref{eq:omnibus_fmui_CCM}, Eqs.~\eqref{eq:omnibus_amui}--\eqref{eq:omnibus_S} yield the results obtained in Refs.~\cite{Newman02_PhysRevLett,Vazquez03_PhysRevE} \footnote{The equivalence between our approach and the one in Refs.~\cite{Newman02_PhysRevLett,Vazquez03_PhysRevE} is obtained through the substitution $u_i\equiv\sum_{j\in\mathcal{N}}P(j|i)[(1-r_jp_{ij})+r_jp_{ij}a_{ij}]$.}.  which allows to study targeted removal of vertices based on their degree in uncorrelated random graphs
%
% Replacing $P(i|j)$ by $\frac{iw_i}{\sum_{j\in\mathcal{N}}jw_j}$ in Eq.~\eqref{eq:omnibus_Pi_CCM} and letting the probability for vertices to exist to depend on their degree then yields the results obtained in Ref.~\cite{Callaway00_PhysRevLett} which allows to study targeted removal of vertices based on their degree in uncorrelated random graphs. 
%
%
%
% =================================================================================================
\subsection{Clustered random graphs} \label{sec:omnibus_example_clustered}
% =================================================================================================
%
We now show how many variants of the CM containing clustered hyperedges (i.e., hyperedges that contain loops) are special cases of the approach presented in this paper.

Since the clustering property is related to the number of triangles found in graphs---hence capturing the idea that \textit{the friend of my friend is also my friend}---it is natural to introduce clustering in graphs through the use of triangles (i.e., three vertices all connected together) \cite{Miller09_PhysRevE,Newman09_PhysRevLett,Serrano06_PhysRevLett}. The simplest clustered graph ensemble then has $|\mathcal{N}|=1$ types of vertices (type 1, $w_1=1$), and $|\mathcal{E}|=2$ types of stubs: two stubs of type A are paired to form undirected edges and three stubs of type B are matched to create triangles. Note that only one stub of type B is required to belong to a triangle even though its contribution amounts to two to the degree of the vertex; stubs can be seen as a membership to an hyperedge. The constraints given by Eq.~\eqref{eq:omnibus_conditions} de facto set the values of $R(\bm{n})=R(n_{A1},n_{B1})$ since $R(2,0)=1-R(0,3)=3\langle k_A \rangle_{P_1} / (3\langle k_A \rangle_{P_1} + 2\langle k_B \rangle_{P_1})$. Assuming that vertices and edges exist with probabilities $r_1$ and $p_{11}$, we obtain from Eqs.~\eqref{eq:omnibus_gnp} and \eqref{eq:omnibus_theta} 
\begin{subequations}
\begin{align}
  \theta_{A1}(\bm{x}) & = (1-r_1p_{11}) + r_1p_{11}x_{A1} \\
  \theta_{B1}(\bm{x}) & = (1-r_1p_{11})^2 + 2r_1p_{11}[1-r_1p_{11}(2-p_{11})]x_{B1} \nonumber \\
                      & + r_1^2p_{11}^2[3-2p_{11}]x_{B1}^2 \ .
\end{align}
\end{subequations}
Using these two functions in Eqs.~\eqref{eq:omnibus_gi}---\eqref{eq:omnibus_K} leads directly to the results obtained in Ref.~\cite{Miller09_PhysRevE,Newman09_PhysRevLett,Shi07_PhysicaA}. Similarly, the results of Ref.~\cite{Ghoshal09_PhysRevE} can be obtained with three types of vertices, $\mathcal{N}=\{1,2,3\}$, and one type of stubs, $\mathcal{E}=\{A\}$, where all hyperedges are triangles containing one vertex of each type [$R(1,1,1)=1$ and $\theta_{Ai}(\bm{x}) = x_{A1}x_{A2}x_{A3}/x_{Ai}$].

Besides triangles, clustering---or any digression from a perfect tree-like structure---has been introduced in random graphs through the inclusion of various categories of hyperedges that involve more than three vertices. For instance, in Ref.~\cite{Gleeson09_PhysRevE,Hackett11_IntJCompSystSci,Newman03b_PhysRevE} clustering is incorporated through fully connected hyperedges, or \textit{cliques}, where vertices or edges exist with given probabilities (i.e., Erd\H{o}s-R\'enyi graphs). In all cases, there is only one type of vertices. We retrieve the model of Ref.~\cite{Newman03b_PhysRevE} by using one type of stubs; $P_1(k_A)$ prescribes the number of cliques to which vertices belong, and $R(n_{A1})$ prescribes the size of cliques (respectively the distributions $r_m$ and $s_n$ in Ref.~\cite{Newman03b_PhysRevE}). In the model considered in Refs.~\cite{Gleeson09_PhysRevE,Hackett11_IntJCompSystSci}, vertices belong to only one clique, but can have many single edges. Since the number of single edges and the size of the clique can be correlated in the original model, there is one type of stubs for each clique size and an additional type for single edges; cliques of size $m$ are formed by matching $m$ stubs of the type assigned to cliques of size $m$. Hence the structure of the graphs is fully prescribed by $P_1(\bm{k})$ whose argument indicates the number of single edges and the size of the clique. The constraints \eqref{eq:omnibus_conditions} then yield
\begin{align}
   R(\bm{n}) = \frac{1}{R''} \sum_{\beta\in\mathcal{E}} (1-\delta_{0,n_{\beta 1}}) \frac{\langle k_\beta \rangle}{n_{\beta 1}} \ ,
 \end{align} 
where $R''=\sum_{\beta\in\mathcal{E}} \langle k_\beta \rangle/n_{\beta 1}$ is the normalization constant. Using these distributions and quantities, our model reproduces the ones presented in Refs.~\cite{Gleeson09_PhysRevE,Hackett11_IntJCompSystSci}. Also, we have used a version of our model that is similar to the one introduced in Ref.~\cite{Hackett11_IntJCompSystSci} to uncover a transition in the effectiveness of immunization strategies \cite{Hebert-Dufresne13_SciRep}.

Finally, two of the most versatile models published to date are also special cases of our model. Reference~\cite{Allard12_JPhysA} is a previous version of the model presented in this paper. The two main differences are that the previous version did not handle site percolation, and that only stubs of same type could be matched to create hyperedges (e.g., forbidding directed edge between vertices of a same type). The model introduced in Ref.~\cite{Karrer10_PhysRevE} can be retrieved from our model with one type of vertices ($|\mathcal{N}|=1$) and with one type of stubs for each \textit{role} that a vertex can play in hyperedges. However, this approach lacks the systematic method offered by Eqs.~\eqref{eq:omnibus_gnp} to solve percolation on each hyperedge beforehand, thereby limiting the number of hyperedges that can effectively be handled analytically.
%
%This model can handle site and bond percolation but requires to solve percolation on each hyperedge beforehand, as in our model. While such calculation consists in a mere enumeration of each possible configuration of existing vertices and edges, it rapidly becomes cumbersome as the number of vertices and edges increases. Equations~\eqref{eq:omnibus_gnp} then offers a systematic method to perform these calculations and therefore further increases the number of different configurations of hyperedges that can be handled analytically.
%
%
%
% =================================================================================================
\subsection{Weak and strong clustering regimes} \label{sec:omnibus_example_clustering}
% =================================================================================================
%
\begin{figure}[tb]
  \centering
  \includegraphics[width = \linewidth]{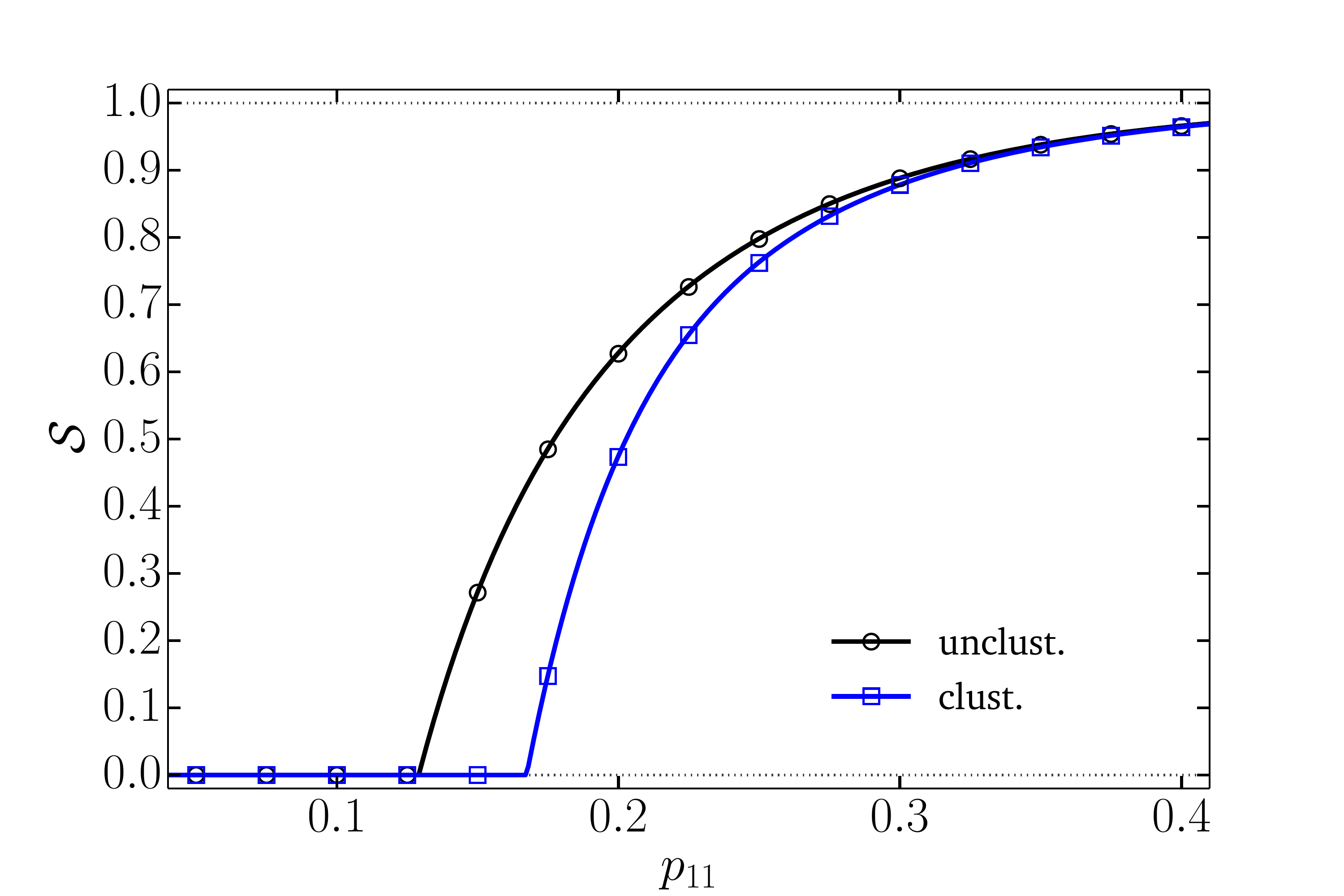} \\
  \caption{\label{fig:omnibus_clustering}(Color online) Comparison of the emergence of the giant component in a clustered graph ensemble that qualifies for the strong clustering regime with its unclustered counterpart. We see that the latter has a larger giant component and a lower percolation threshold. Details on the graphs used are given in Sec.~\ref{sec:omnibus_example_clustering}.}
\end{figure}
We now use our model to test a conjecture regarding the effect of clustering (e.g., triangles) on bond percolation. References~\cite{Serrano06_PhysRevLett,Serrano06b_PhysRevE} proposed that clustering has opposite effects on the bond percolation threshold and on the size of giant component depending of the density of triangles in a graph. This density is measured through the degree dependent clustering coefficient $\bar{c}(k)$: the probability that two neighbors of a vertex of degree $k$ are also neighbors (i.e., they complete the triangle). The conjecture states that the \textit{weak} clustering regime $\bar{c}(k)<(1-k)^{-1}$ leads to a higher percolation threshold and to a smaller giant component than in an equivalent unclustered graph. Contrariwise, \textit{strong} clustering, defined as $\bar{c}(k)>(1-k)^{-1}$, leads to a lower percolation threshold and to a larger giant component than in an equivalent unclustered graph.

Let us consider the following graph ensemble in which there are two types of vertices $\mathcal{N}=\{1,2\}$ and three types of edges $\mathcal{E}=\{A,B,C\}$. Every vertex of type 1 has one stub of type A and one stub of type B, while each vertex of type 2 has one stub of type B and one stub of type C. In other words, we set $P_i(\bm{k})=P_i(k_A,k_B,k_C)$ as $P_1(1,1,0)=P_2(0,1,1)=1.0$. Hyperedges are formed by matching either 4 stubs of type A, 4 stubs of type B (two stemming from vertices of type 1 and two from vertices of type 2), or 8 stubs of type C; vertices are all connected to one another in every hyperedges. The constraints given by Eq.~\eqref{eq:omnibus_conditions} imply that $w_1=w_2$ and that $R(\bm{n})=R(n_{A1},n_{B1},n_{C1},n_{A2},n_{B2},n_{C2})$ follows the relation $2R(4,0,0,0,0,0)=R(0,2,0,0,2,0)=4R(0,0,0,0,0,8)=4/7$. Vertices of type 1 all have a degree equal to 6, and vertices of type 2 all have a degree equal to 10. Consequently, we see that $\bar{c}(6) = \frac{6}{15}>\frac{1}{5}$ and $\bar{c}(10)=\frac{24}{45}>\frac{1}{9}$, which implies that this graph ensemble qualifies for the strong regime.

To isolate the effect of clustering on bond percolation, we compare the results obtained for the graph ensemble described above with the ones obtained with an equivalent unclustered version \cite{Miller09_PhysRevE}. This equivalent graph ensemble possesses identical correlations, but hyperedges are broken into individual independent edges instead (e.g., each vertex in an hyperedge containing $n$ vertices now have $n-1$ independent edges). The behavior of its giant component is obtained as in Sec.~\ref{sec:omnibus_example_correlated} with $P_1(4,2,0)=P_2(0,2,8)=1$.

Figure~\ref{fig:omnibus_clustering} compares the behavior of the giant component in both ensembles when edges exist with probability $p_{11}$. We conclude that although the clustered graph ensemble qualifies for the strong regime, the behavior observed is the one of the weak regime: higher percolation threshold and smaller giant component than for the equivalent unclustered graph. This behavior can be understood in terms of branching factors. The unclustered graphs have a tree-like structure and therefore maximize the number of \textit{new} vertices encountered while navigating the graph: every edge leads to a new vertex. The redundancy caused by clustering means that not all edges lead to a new vertex in the clustered graphs, which reduces the average number of vertices that can be reached from any given vertex. Hence a larger number of edges must be present for a giant component to appear (e.g., larger threshold), and this component will be smaller as many edges will be \textit{wasted} by leading to vertices previously reached. 

This counterexample suggests that the criterion on $\bar{c}(k)$ could be a necessary condition for a strong clustering regime but that it is not a sufficient one. The explanation in terms of branching factors alongside the results in Refs.~\cite{Gleeson10_PhysRevE,Kiss08_PhysRevE,Miller09_PhysRevE} point toward the conclusion that the effect of clustering on random graphs with an \textit{underlying} tree-like structure is best described by the weak clustering regime. Indeed, Ref.~\cite{Colomer-de-Simon2014e} has recently shown that strong clustering may induce a double continuous phase transition which conciliates the conjectured antagonistic effects of weak and strong clustering, and whose effects are in line with our conclusion.
%
%
%
% =================================================================================================
\subsection{Bijection between site and bond percolation thresholds}
% =================================================================================================
%
From Eqs.~\eqref{eq:omnibus_fmui}, we see that the elements of the Jacobian matrix $\mathbf{J}$ used to determine the point at which the giant component appears have the general form
\begin{align} \label{eq:omnibus_Jacobian_elements}
  \frac{\partial f_{\mu i}(\bm{1})}{\partial x_{\nu j}} & = \sum_{\alpha\in\mathcal{E}} \frac{\langle k_\mu(k_\alpha-\delta_{\mu\alpha})\rangle_{P_i}}{\langle k_\mu \rangle_{P_i}} \frac{\partial \theta_{i\alpha}(\bm{1})}{\partial x_{\nu j}} \ ,
\end{align}
for every $i,j\in\mathcal{N}$ and $\mu,\nu\in\mathcal{E}$. These terms are in fact branching factors: each element is the average number of pairs $(\nu,j)$ that are present in the neighborhood of a pair $(\mu,i)$. More precisely, the first term corresponds to the average number of stubs of type $\alpha$ that a vertex of type $i$ has if it has been reached from one of its stubs of type $\mu$ (this stub is excluded from the count if $\alpha=\mu$). The second term is the average number of pairs $(\nu,j)$ that can be reached in hyperedges accessed via a stub of type $\alpha$ of a vertex of type $i$. The value of these latter terms depends on the structure of hyperedges (i.e., rules $\mathcal{R}$) and on the probabilities for vertices and edges to exist (i.e., $\{r_j\}_{j\in\mathcal{N}}$ and $\{p_{jk}\}_{j,k\in\mathcal{N}}$).

Let us assume that all hyperedges have the same structure; vertices of different types may be involved in a nontrivial manner as long as all hyperedges have the same shape (e.g., they all are triangles). We also suppose that vertices and edges exist with probabilities that are independent of their type, that is $r_i=r$ and $p_{ij}=p$ for all $i,j\in\mathcal{N}$. In such case, every nonzero elements of the Jacobian matrix, $\frac{\partial \theta_{i\alpha}(\bm{1})}{\partial x_{\nu j}}$, is a polynomial in $r$ and $p$, $h(r,p)$, and is independent of $i$, $j$, $\alpha$ and $\nu$. Consequently the dependency in $r$ and $p$ can be factored out of the Jacobian matrix
\begin{align} \label{eq:omnibus_bijthres_facto}
  \mathbf{J}=h(r,p)\mathbf{J}' \ .
\end{align}
Since the giant component appears when $\lambda_{\mathrm{max}}(\mathbf{J}) = h(r,p) \lambda_{\mathrm{max}}(\mathbf{J}') = 1$, the points $(r',p')$ at which the phase transition occurs all belong to the critical surface
\begin{align}
  h(r',p') = \frac{1}{\lambda_{\mathrm{max}}(\mathbf{J}')} \ . 
\end{align}
Whenever the Jacobian matrix can be written like in Eq.~\eqref{eq:omnibus_bijthres_facto}, any given point $(r_1,p_1)$ at which a graph ensemble is known to percolate can be related to any other critical point $(r_2,p_2)$ through $h(r_1,p_1)=h(r_2,p_2)$. For instance, this relation leads to a direct bijection between the thresholds of pure site percolation $(r_c,1)$ and pure bond percolation $(1,p_c)$ through $h(r_c,1)=h(1,p_c)$. Additionally, $h(r,p)=rp$ for unclustered correlated random graphs, and the fact that $r$ and $p$ only appear as $rp$ in Eqs.~\eqref{eq:omnibus_theta_correlated} implies that site and bond percolation are equivalent for this random graph ensemble.
%
%
%
%
%
%
% =================================================================================================
\section{Interdependent random graphs} \label{sec:omnibus_discontinuous}
% =================================================================================================
%
In this last section, we briefly show how our approach can be adapted to model interdependent graphs through a redefinition of Eqs.~\eqref{eq:omnibus_amui}--\eqref{eq:omnibus_S}, and use the resulting formalism to investigate the emergence of an extensive component on interdependent clustered random graphs.

To lighten the description, we consider the case of two interdependent graphs, graph A and graph B, without loss of generality (guidelines for a straightforward generalization to an arbitrary number of graphs are given in Ref.~\cite{Son12_EPL}). We assume that every edge in each graph is undirected such that there is no global asymmetry: the probability that a randomly chosen vertex leads to the extensive component is equal to the relative size of the extensive component (i.e., $\mathcal{P}=\mathcal{S}$). Furthermore, we consider that the pgfs $g_i(\bm{x})$ and $f_{\mu i}(\bm{x})$ and all other related quantities defined in the previous sections (i.e., $\{w_i\}$, $\mathcal{R}$, $\mathcal{N}$, $\{r_i\}$, \ldots) are known for both graphs and are identified with the superscript A or B. Both graphs contain the same number of vertices which tends to infinity: $N^A=N^B=N\rightarrow\infty$.

The change in the nature of the transition (i.e., from continuous to discontinuous) originates from the existence of \textit{dependencies} between vertices of the two graphs. Again, to lighten the description, we consider the case in which each vertex has either one \textit{twin vertex} on which it depends, or none. To specify the dependencies between vertices, we define $q_{iv}^{AB}$ as the probability that a vertex of type $i$ in graph A has a twin vertex of type $v$ in graph B. Note that allowing graphs to be partially dependent---not all vertices have a twin vertex---implies that
\begin{align}
  \sum_{v\in\mathcal{N}^B}q_{iv}^{AB} \equiv 1 - \bar{q}_i^{AB} \leq 1 \ ,   
\end{align}
for each $i\in\mathcal{N}^A$, and where the sum is over the types of the vertices in graph B. Therefore, a fraction $\bar{q}_i^{AB}$ of the vertices of type $i$ in graph A do not have a twin vertex in graph B. A similar set of probabilities, $\{q_{ju}^{BA}\}_{j,u}$ with $j\in\mathcal{N}^B$ and $u\in\mathcal{N}^A$, is defined to specify the dependencies of vertices in graph B. Moreover, we add the constraint that the dependency between two vertices must be reciprocal unless a vertex's twin has no dependency whatsoever. In other words, if vertex $n^A$ in graph A depends on vertex $n^B$ in graph B, then either vertex $n^B$ depends on vertex $n^A$ as well or it depends on no vertex at all. In the latter case, vertex $n^A$ is the only vertex in graph A that can depend on vertex $n^B$. This constrains the two probability sets, $\{q_{iv}^{AB}\}_{i,v}$ and $\{q_{ju}^{BA}\}_{j,u}$, as there must be enough ``independent'' vertices in graph A (graph B) to account for the vertices in graph B (graph A) whose dependency is not reciprocal. Mathematically, these conditions can be written as
\begin{align} \label{eq:omnibus_cond_AB} 
  \sum_{i\in\mathcal{N}^A} \mathrm{max}\Big\{N^Aw_i^Aq_{iv}^{AB}-N^Bw_{v}^Bq_{vi}^{BA},0\Big\} & \leq N^Bw_v^B \bar{q}_v^{BA}
\end{align}
for each $u\in\mathcal{N}^A$ and $v\in\mathcal{N}^B$. A similar expression in which the superscripts A and B are swapped must also hold. In the expression above, $N^Aw_i^Aq_{iv}^{AB}$ corresponds to the number of vertices of type $i$ in graph A that depend on a vertex of type $v$ in graph B, and $N^Bw_v^B \bar{q}_v^{BA}$ is the number of vertices of type $v$ in graph B that have no dependency.

A discontinuous phase transition is associated with the emergence of an extensive \textit{functional} component: an extensive component composed solely of vertices with no dependency or whose twin vertex is part of the extensive functional component in its respective graph. To compute the size of the extensive functional component, we define $a_{\mu i}^{A}$ as the probability that a vertex of type $i$ reached from one of its stubs of type $\mu$ in graph A \textit{does not} lead to the functional extensive component in graph A. Similar probabilities are defined for the other types of vertices and stubs in graph A and in graph B. Following the locally tree-like structure argument of Sec.~\ref{sec:omnibus_extensive_component}, we now derive a set of self-consistent equations similar to Eqs.~\eqref{eq:omnibus_amui} for these probabilities.

Let us consider the case of a vertex of type $i$ in graph A reached via one of its stubs of type $\mu$. By definition, this vertex belongs to the extensive functional component in graph A with probability $1-a_{\mu i}^A$. Only two scenarios can lead to this situation. The first one consists in the vertex being part of the extensive functional component in graph A and having no twin vertex. This happens with probability $[1 - f_{\mu i}^A(\bm{a^A})] \bar{q}_i^{AB}$. The second scenario consists in the vertex being part of the extensive functional component in graph A and having a twin vertex that is part of the extensive functional component in its own graph. This scenario happens with probability
\begin{align}
  \Big[1 - f_{\mu i}^A(\bm{a^A})\Big] \sum_{v\in\mathcal{N}^B}q_{iv}^{AB}r_v^B\big[1-g_v^B(\bm{a^B})\big] \ ,
\end{align}
where $q_{iv}^{AB}$ is the probability that the twin vertex is of type $v$, $r_v^B$ is the probability that it exists (i.e., it has not been removed), and $1-g_v^B(\bm{a^B})$ is the probability that it belongs to the extensive functional component in graph B. Summing these two scenarios yields
\begin{multline} \label{eq:omnibus_a_AB}
  a_{\mu i}^{A} = 1 - \Big[1 - f_{\mu i}^A(\bm{a^A})\Big] \Big[ \bar{q}_i^{AB} \\ + \sum_{v\in\mathcal{N}^B}q_{iv}^{AB}r_v^B\big[1-g_v^B(\bm{a^B})\big] \Big] \ ,
\end{multline}%
which must hold for every $i\in\mathcal{N}^A$ and $\mu\in\mathcal{E}^A$, as well as for graph B (i.e., simply swap the superscripts A and B in the last expression). Having solved Eqs.~\eqref{eq:omnibus_a_AB} for the probabilities $\bm{a^A} \equiv \{a_{\mu i}^A\}$ and $\bm{a^B} \equiv \{a_{\nu j}^B\}$, the probability that a randomly chosen vertex of type $i$ in graph A belongs to the functional component is
\begin{multline}
  \mathcal{S}_i^A = \Big[1 - g_{i}^A(\bm{a^A})\Big] \Big[ \bar{q}_i^{AB} \\ + \sum_{v\in\mathcal{N}^B}q_{iv}^{AB}r_v^B\big[1-g_v^B(\bm{a^B})\big] \Big] \ ,
\end{multline}
which is similar to the calculation of $\mathcal{S}_i$ in Sec.~\ref{sec:omnibus_extensive_component}, but in this case the probability that a vertex of type $i$ belongs to the functional extensive component, $1 - g_{i}^A(\bm{a^A})$, is weighted by the probability that its twin vertex, if any, belongs to the extensive functional component as well. Averaging over the fraction of existing vertices of each type (e.g., a fraction $r_i^A w_i^A$ of vertices in graph A corresponds to vertices of type $i$ that have not been removed), we finally obtain the size of the extensive functional component in graph A
\begin{align} \label{eq:omnibus_S_AB}
  \mathcal{S}^A & = \sum_{i\in\mathcal{N}^A}\frac{r_i^Aw_i^A}{\sum_{j\in\mathcal{N}^A}r_j^Aw_j^A}\mathcal{S}_i^A \ .
\end{align}
Similar equations for vertices in graph B are obtained by swapping the superscripts A and B in the last two expressions. As for the quantities $\mathcal{P}$ and $\mathcal{S}$ defined previously, the fraction $\mathcal{S}^A$ (and $\mathcal{S}^B$) is relative to the number of existing vertices (i.e., vertices that have not been removed). Note also that  Eqs.~\eqref{eq:omnibus_amui} are retrieved from Eqs.~\eqref{eq:omnibus_a_AB} if there are no dependencies. However, contrariwise to Eqs.~\eqref{eq:omnibus_amui} and Eqs.~\eqref{eq:omnibus_bara_mui}, the right-hand side of Eqs.~\eqref{eq:omnibus_a_AB} does not necessarily correspond to monotonously increasing functions (some coefficients in the polynomials are negative). This implies that, although the point $\bm{a^A} \oplus \bm{a^B}=\bm{1}$ is still a solution, another solution in the hypercube $[0,1]^{|\mathcal{N}^A|\times|\mathcal{E}^A|+|\mathcal{N}^B|\times|\mathcal{E}^B|}$ corresponding to the presence of an extensive functional component may not appear continuously from $\bm{a^A} \oplus \bm{a^B}=\bm{1}$ through a transcritical bifurcation as in Sec.~\ref{sec:omnibus_extensive_component}. Hence the values of $\mathcal{S}^A$ and $\mathcal{S}^B$ \textit{jump} abruptly from zero to a finite value in [0,1] which corresponds to a discontinuous phase transition.

\begin{figure}[tb]
  \centering
  \includegraphics[width = \linewidth]{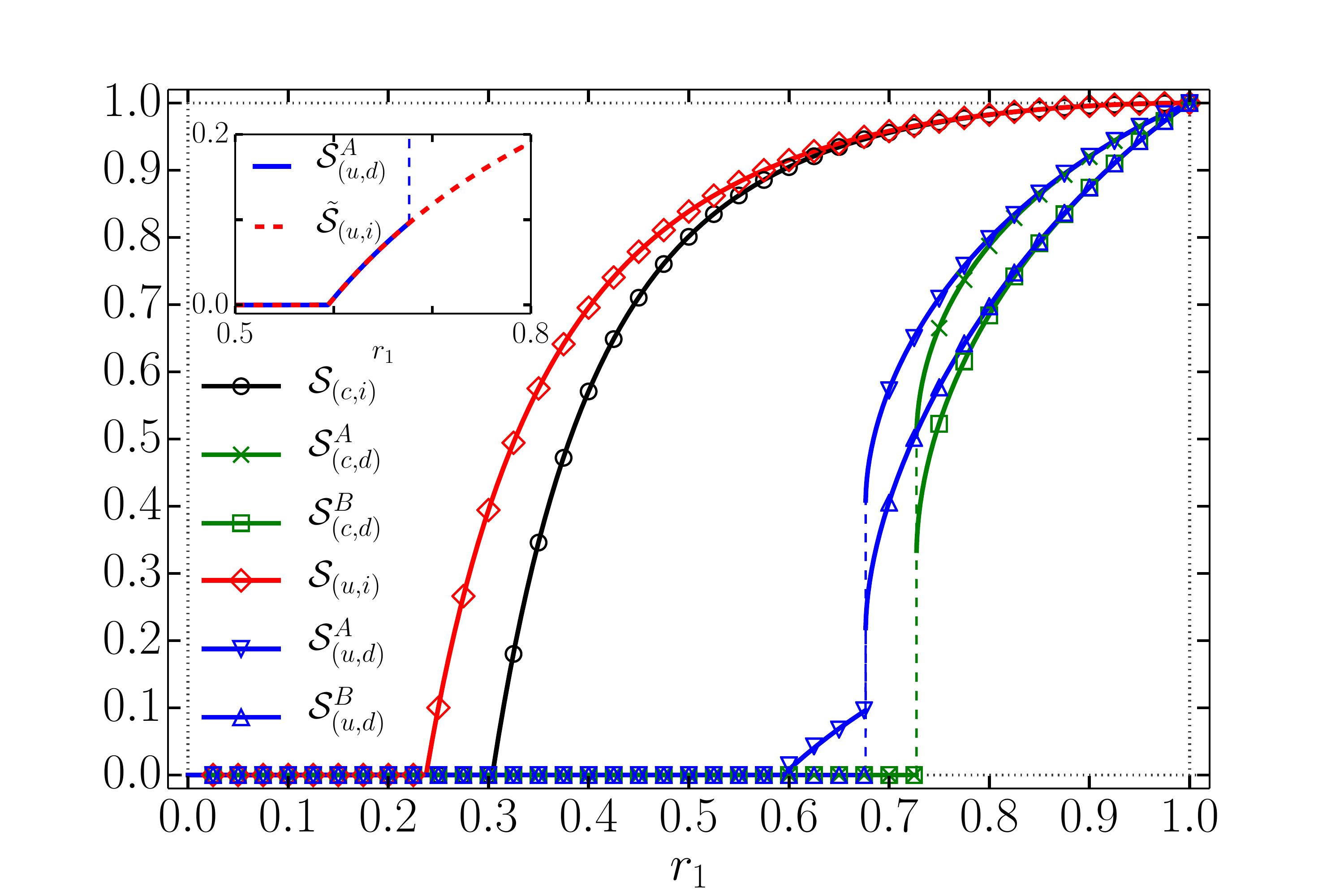} \\
  \caption{\label{fig:omnibus_discontinuous}(Color online) Comparison of the size of the extensive (functional) components as a function of the vertex existence probability $r_1$ in four related graph ensembles. The details of each graph ensemble are given in the main text of Sec.~\ref{sec:omnibus_discontinuous}. The curves $\mathcal{S}_{(c,i)}$ and $\mathcal{S}_{(u,i)}$ were obtained with Eq.~\eqref{eq:omnibus_S} and the curves $\mathcal{S}^A_{(c,d)}$, $\mathcal{S}^B_{(c,d)}$, $\mathcal{S}^A_{(u,d)}$ and $\mathcal{S}^B_{(u,d)}$ were obtained with Eq.~\eqref{eq:omnibus_S_AB}. Symbols show the results of numerical simulations on these graph ensembles with $N=10^6$ vertices. (inset) Comparison between the curve $\mathcal{S}^A_{(u,d)}$ and the curve $\mathcal{S}_{(u,i)}$ rescaled according to $\tilde{S}_{(u,i)}[r_1]= \bar{q}_{1}^{AB} S_{(u,i)}[r_1\bar{q}_{1}^{AB}]$ (the dependence to $r_1$ in shown in brackets).}
\end{figure}

To illustrate this behavior, we investigate the emergence of extensive components on interdependent clustered random graphs. To do so, we consider the edge-triangle clustered graph ensemble presented in Sec.~\ref{sec:omnibus_example_clustered} with the joint degree distribution $P_1(0,3)=P_1(2,1)=2P_1(2,0)=4/10$ and $p_{11}=1.0$. Notice that this joint degree distribution forces assortative mixing since high and low degree vertices tend to be segregated. The size of the extensive component in this isolated random graph ensemble is given as a function of the vertex existence probability $r_1$ in Fig.~\ref{fig:omnibus_discontinuous} (black curve labeled $\mathcal{S}_{(c,i)}$).

To illustrate the impact of interdependence on the phase transition, we consider the case of two identical partially dependent edge-triangle clustered graphs with $q_{11}^{AB}=0.6$ and $q_{11}^{BA}=1.0$. In other words, only 60\% of the vertices in graph A depend on a vertex in graph B whereas every vertex in graph B has a twin vertex. The green curves labeled $\mathcal{S}^A_{(c,d)}$ and $\mathcal{S}^B_{(c,d)}$ in Fig.~\ref{fig:omnibus_discontinuous} show the size of the extensive functional component in graph A and graph B as a function of $r_1$. We see that the extensive functional components indeed emerge through a discontinuous phase transition, unlike the extensive component in the isolated clustered graphs that emerges continuously. We also see that the extensive functional component in graph A is always bigger than the one on graph B since vertices with no dependency are more likely to be in the extensive functional component than vertices with a dependency. Indeed, a vertex in graph A that has no dependency belongs to the extensive functional component with probability $[1-g_1^A(\bm{a^A})]$ which is clearly greater than $[1-g_1^A(\bm{a^A})]r_1^B[1-g_1^B(\bm{a^B})]$ for a vertex with a dependency since $0\leq r_1^B[1-g_1^B(\bm{a^B})] \leq 1$ (the second equality holds only when there is no extensive component and $r_1^B=1$). Figure~\ref{fig:omnibus_discontinuous} also shows that the size of the extensive functional component in the interdependent graphs is bounded by the size of the giant component in the corresponding isolated graphs. Again, this is expected since being part of the extensive component is a \textit{sine qua non} condition for being part of the extensive functional component.

As in Sec.~\ref{sec:omnibus_example_clustering}, we consider the unclustered version of the edge-triangle random graph ensemble---in which triangles are broken into two independent single edges---to isolate the impact of clustering. In order to preserve the correlations present in the clustered dependent graph ensemble, two types of stubs are used to distinguish the original single edges from the single edges due to the broken triangles. As expected, the extensive component in this isolated unclustered graph ensemble appears at a lower value of $r_1$ (red curve labeled $\mathcal{S}_{(u,i)}$ in Fig.~\ref{fig:omnibus_discontinuous}) than for the isolated clustered random graph ensemble (see Sec.~\ref{sec:omnibus_example_clustering}). The same conclusion holds for the extensive functional component in the unclustered version of the two partially dependent graphs described in the last paragraph (blue curves labeled $\mathcal{S}^A_{(u,d)}$ and $\mathcal{S}^B_{(u,d)}$ in Fig.~\ref{fig:omnibus_discontinuous}). Comparing the size of the extensive functional component in the clustered and unclustered versions of these interdependent graphs also suggests that clustering increases the jump size at the transition.

One interesting observation from Fig.~\ref{fig:omnibus_discontinuous} is that graph A in the interdependent unclustered graph ensemble (curve labeled $\mathcal{S}^A_{(u,d)}$) successively undergoes two phase transitions: a continuous then a discontinuous one. While the discontinuous transition is caused by the interdependency with graph B, the continuous one is due to the fact that the 40\% of vertices in graph A that have no dependency are able to form an extensive component before the discontinuous phase transition occurs. From Fig.~\ref{fig:omnibus_discontinuous}, we see that the continuous phase transition happens at $r_1\simeq0.23$ in the isolated unclustered graph ensemble, while the phase transitions occur at $r_1\simeq0.58$ (continuous) and at $r_1\simeq0.68$ (discontinuous) in the interdependent unclustered graph ensemble. Below $r_1\simeq0.68$, the vertices in graph A that depend on a vertex in graph B can be effectively considered as removed since there is no extensive functional component in graph B (i.e., they cannot be part of an extensive component). Hence we expect graph A to behave as its isolated graph ensemble counterpart in which an effective fraction $1-r_1\bar{q}_{1}^{AB}$ of its vertices have been removed. This is indeed confirmed in the inset of Fig.~\ref{fig:omnibus_discontinuous}. In fact, successive phase transitions should occur whenever independent vertices are able to form an extensive component \textit{before} the extensive functional component emerges. In other words, we observe a double phase transition whenever the rescaled value at which the continuous phase transition happens in the isolated graph ensemble is below the value at which the discontinuous phase transition occurs in the interdependent graph ensemble.
%
%
%
%
%
% ====================================================================================================
\section{Conclusion} \label{sec:omnibus_conclusion}
% ====================================================================================================
%
Building upon our previous works \cite{Allard12_JPhysA,Allard12_EPL,Allard09_PhysRevE,Hebert-Dufresne13_SciRep,Hebert-Dufresne13_PhysRevE}, we have presented a unifying conceptual framework that offers a comprehensive mathematical description of a wide variety of structural properties found in graphs extracted from real complex systems (e.g., correlations, segregation, clustering of various forms). The generality of the formalism resides on a multitype perspective for a precise prescription on how vertices are connected to one another, and on a set of iterative equations for the solution of the distribution of the size of components in small arbitrary graphs. Interestingly, these iterative equations are by themselves a valuable addition to graph theoretical methodology. In fact, besides being a cornerstone of our formalism, allowing a mapping of hyperedges unto an effective tree-like structure, they also have potential applications in the theoretical description of fragmentation processes and of percolation on lattices \cite{Desesquelles11_PhysLettB,Desesquelles13_JPhysConfSer,Reynolds80_PhysRevB,Scullard08_PhysRevLett} (see Ref.~\cite{Allard12_EPL} for further details).

Our approach leads to the definition of a very general random graph ensemble for which site and/or bond percolation can be solved exactly using probability generating functions in the infinite size limit (e.g., size of the giant component, percolation threshold, distribution of the size of small components). We have shown that this random graph ensemble encompasses most random graph models published until now and can incorporate structural properties not yet included in a theoretical framework. This versatility makes it a perfect theoretical laboratory to investigate the role of specific local structural properties on the global connectivity of the graphs. We have illustrated this point by implementing our method to provide a counterexample to a conjecture \cite{Serrano06_PhysRevLett} on the effect of clustering on the size of the giant component and on the percolation threshold.

Our formalism is also naturally equipped for the modeling of interdependent graphs whose most striking feature is the emergence of the extensive component via a discontinuous phase transition. We have provided a specific implementation for this application that demonstrates how a graph can successively undergo a continuous then a discontinuous phase transition, and how clustering increases the amplitude of discontinuity at the transition.

By offering one of the most comprehensive mathematical description of percolation on random graphs, we believe that the present work will contribute to a better understanding of the interplay between local structural properties and the global connectivity of graphs. Moreover, our approach can easily accommodate other types of dynamics for which the pgf technique has already proven to be useful \cite{Allard2014,Hackett11_PhysRevE,Watts2002a}. We are hopeful that several extensions (different dynamics and/or percolation models) will shed further light on the role of structure in the behavior of complex systems. We put forward that some of the tools to perform these studies are now available.
\begin{acknowledgments}
  We acknowledge support from the Instituts de recherche en sant\'e du Canada, the Conseil de recherches en sciences naturelles et en g\'enie du Canada and the Fonds de recherche du Qu\'ebec -- Nature et technologies.
\end{acknowledgments}
%
%
%
%
%
% ====================================================================================================
% Bibliography
% ====================================================================================================
%
% \bibliography{/home/allard/Documents/biblio/library}
%merlin.mbs apsrev4-1.bst 2010-07-25 4.21a (PWD, AO, DPC) hacked
%Control: key (0)
%Control: author (0) dotless jnrlst
%Control: editor formatted (1) identically to author
%Control: production of article title (0) allowed
%Control: page (1) range
%Control: year (0) verbatim
%Control: production of eprint (0) enabled
%

%
%
%
\end{document}